\documentclass[conference,compsoc]{IEEEtran}
\IEEEoverridecommandlockouts

\ifCLASSOPTIONcompsoc
  \usepackage[nocompress]{cite}
\else
  \usepackage{cite}
\fi

\ifCLASSINFOpdf
\else
\fi

\usepackage{xcolor}
\usepackage{float}
\usepackage{tikz}
\usepackage{amsmath}
\usepackage{url}
\usepackage{xspace}
\usepackage{comment}

\definecolor{customblue}{HTML}{334A8D}
\usepackage[colorlinks=true,linkcolor=customblue,citecolor=customblue,urlcolor=customblue]{hyperref}
\usepackage{listings}
\makeatletter
\let\orig@lstKV@SwitchCases\lstKV@SwitchCases
\renewcommand*\lstKV@SwitchCases[3]{}
\makeatother
\usepackage{lstlinebgrd}
\makeatletter
\let\lstKV@SwitchCases\orig@lstKV@SwitchCases
\lst@Key{numbers}{none}{%
  \def\lst@PlaceNumber{\lst@linebgrd}%
  \lstKV@SwitchCases{#1}{%
    none:\def\lst@PlaceNumber{\lst@linebgrd}\\%
    left:\def\lst@PlaceNumber{\llap{\normalfont
      \lst@numberstyle{\thelstnumber}\kern\lst@numbersep}\lst@linebgrd}\\%
    right:\def\lst@PlaceNumber{\rlap{\normalfont
      \kern\linewidth \kern\lst@numbersep
      \lst@numberstyle{\thelstnumber}}\lst@linebgrd}%
  }{\PackageError{Listings}{Numbers #1 unknown}\@ehc}}
\makeatother
\usepackage{amssymb}
\usepackage{empheq}
\usepackage{booktabs}
\usepackage{multirow}
\usepackage{multicol}
\usepackage[ruled,vlined,resetcount,linesnumbered]{algorithm2e}
\usepackage{amsthm}
\usepackage{colortbl}
\usepackage{enumitem}
\usepackage[binary-units=true,detect-all]{siunitx}

\usepackage{pgfplots}
\usepackage{pifont}

\usepackage{balance}
\usepackage[capitalise,noabbrev]{cleveref}

\usepackage[scaled=0.95]{inconsolata}

\theoremstyle{definition}

\newcommand{\tool}{\textsc{Code-Augur}\xspace}
\newcommand{\atlantis}{\textsc{Atlantis}\xspace}
\newcommand{\claude}{\textsc{Claude Code}\xspace}

\newcommand{\myparagraph}[1]{{\textbf{#1}}}

\usepackage[T1]{fontenc}

\newcommand{\bugnum}{22\xspace}
\newcommand{\fixedbugnum}{16\xspace}

\newcommand{\bugfold}{$34\%$--$370\%$\xspace}

\definecolor{lightblue}{RGB}{70, 130, 180}

\definecolor{workfloworange}{HTML}{FFEED9}
\definecolor{workflowblue}{HTML}{D9E7FD}
\definecolor{workflowgray}{HTML}{F2F2F2}
\definecolor{workflowpink}{HTML}{FBDCE3}

\definecolor{claudecol}{HTML}{D97757}
\definecolor{deepseekcol}{HTML}{4D6BFE}
\definecolor{claudebase}{RGB}{255,243,217}
\definecolor{deepseekbase}{RGB}{218,232,252}
\colorlet{claudebg}{claudebase!55}
\colorlet{deepseekbg}{deepseekbase!42}
\newcolumntype{O}{>{\columncolor{claudebg}}r}
\newcolumntype{B}{>{\columncolor{deepseekbg}}r}
\newcommand{\claudelogo}{\raisebox{-0.18em}{\includegraphics[height=1.0em]{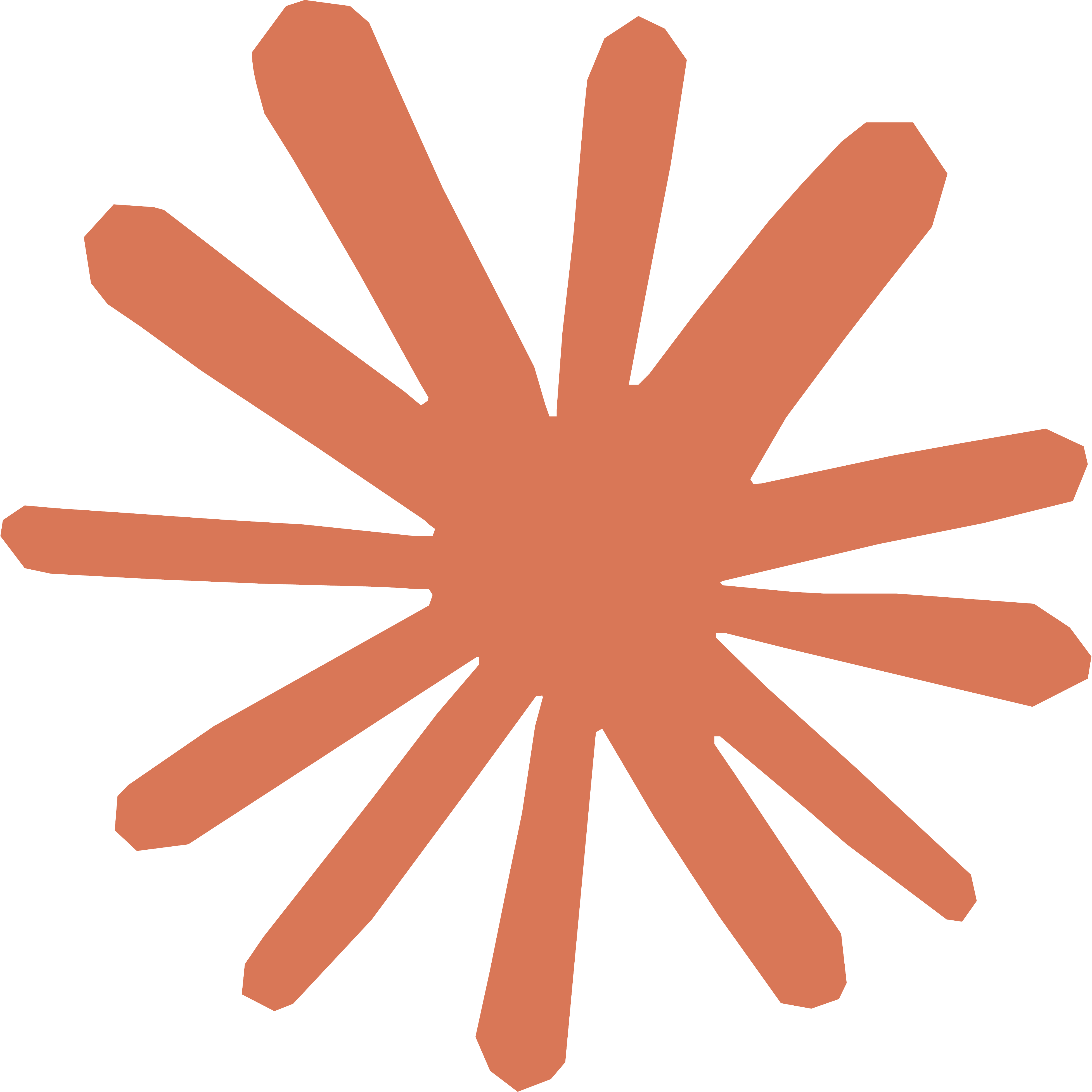}}}
\newcommand{\deepseeklogo}{\raisebox{-0.30em}{\includegraphics[height=1.3em]{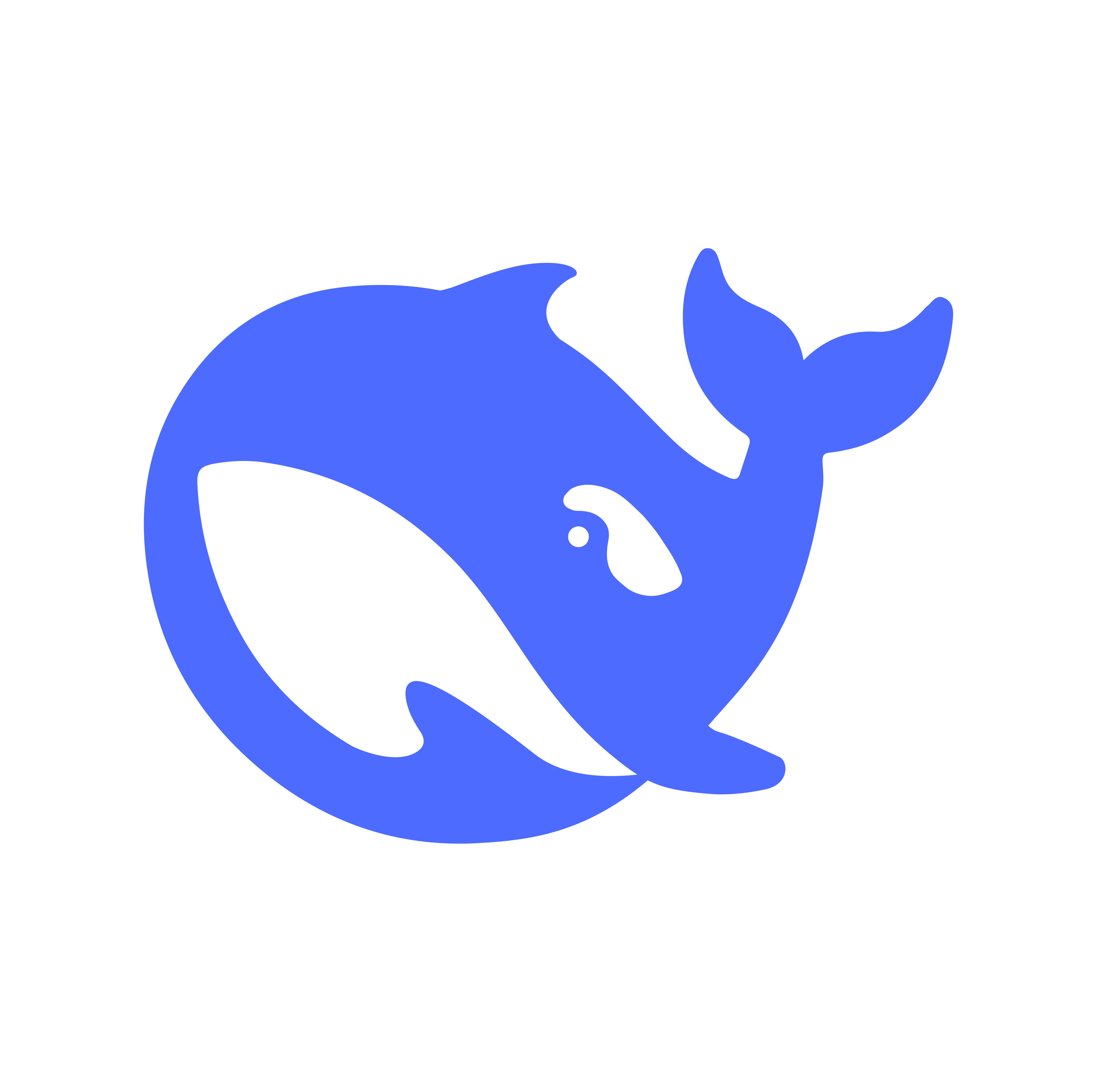}}}
\newcommand{\impr}[1]{\,{\footnotesize(+#1\%)}}
\newcommand{\resultsnote}{%
  \par\smallskip
  {\footnotesize
  \begin{itemize}[nosep,leftmargin=1.4em]
    \item \textbf{Existing}: known bugs reproduced (``\emph{x/total}'').
   \textbf{New}: newly discovered bugs at the benchmarked version.
   \textbf{Total~(Existing+New)}: the two combined.
    \item \textbf{Parenthesized \%}: \tool's relative improvement over that baseline under the same backing model, computed over the subjects where both ran successfully.
  \end{itemize}}%
}

\newcommand{\hlbox}[2]{{\setlength{\fboxsep}{1pt}\colorbox{#1}{\rule[-0.25ex]{0pt}{1.55ex}#2}}}

\makeatletter
\newcommand*{\lstlinelabel}[1]{%
  \protected@edef\@currentlabel{\thelstnumber}%
  \label{#1}%
}
\makeatother

\definecolor{lightgray}{rgb}{0.9,0.9,0.9}
\definecolor{lightgray2}{rgb}{0.96,0.96,0.96}
\definecolor{hdmlblue}{RGB}{141, 180, 226}
\definecolor{babyblueeyes}{rgb}{0.63, 0.79, 0.95}
\definecolor{beaublue}{rgb}{0.74, 0.83, 0.9}
\definecolor{softblue}{rgb}{0.3,0.4,0.6}
\definecolor{codeblue}{rgb}{0.0, 0.0, 0.8}
\definecolor{codegreen}{rgb}{0,0.6,0}

\definecolor{colFmt}{RGB}{0,128,128}
\definecolor{colChan}{RGB}{198,83,0}
\definecolor{colCS}{RGB}{0,112,192}
\definecolor{colModel}{RGB}{125,45,145}

\newcommand{\Cfmt}{\mathbb{C}_{\texttt{PixelFormat}}}
\newcommand{\Ccs}{\mathbb{C}_{\texttt{ColorSpace}}}

\makeatletter
\newcommand{\beh@parens}{\@ifnextchar\bgroup{\beh@wrap}{}}
\newcommand{\beh@wrap}[1]{(#1)}
\newcommand{\beh}[1]{\Pi_{#1}\beh@parens}
\newcommand{\behm}[1]{\hat{\Pi}_{#1}\beh@parens}
\newcommand{\vbeh}[1]{\Pi^{\viol}_{#1}\beh@parens}
\newcommand{\vbehm}[1]{\hat{\Pi}^{\viol}_{#1}\beh@parens}
\makeatother
\newcommand{\isus}{i^{*}}
\newcommand{\viol}{\text{\textsc{err}}}

\SetKwInOut{Input}{Input}
\SetKwInOut{Output}{Output}
\SetKwProg{myalg}{Algorithm}{}{}
\SetKwProg{myproc}{Procedure}{}{}

\lstdefinestyle{cstyle}{
  language=C,
  basicstyle=\ttfamily\scriptsize,
  keywordstyle=\color{codeblue}\bfseries,
  commentstyle=\color{codegreen}\itshape,
  stringstyle=\color{codegreen},
  numbers=none,
  showstringspaces=false,
  breaklines=true,
  columns=fullflexible,
  escapeinside={(*@}{@*)},
  aboveskip=4pt,
  belowskip=2pt,
  captionpos=b,
  xleftmargin=4pt,
  morekeywords=[2]{Format,fmt},                    keywordstyle=[2]\color{colFmt},
  morekeywords=[3]{channels},                      keywordstyle=[3]\color{colChan},
  morekeywords=[4]{ColorSpace,cs},      keywordstyle=[4]\color{colCS},
  morekeywords=[5]{colormodel,PT_ANY},keywordstyle=[5]\color{colModel},
}

\lstdefinestyle{tmstyle}{
  basicstyle=\ttfamily\scriptsize,
  backgroundcolor=\color{workflowblue!22},
  frame=single,
  rulecolor=\color{workflowblue!80!black},
  framerule=0.4pt,
  numbers=none,
  showstringspaces=false,
  breaklines=true,
  columns=fullflexible,
  aboveskip=7pt,
  belowskip=4pt,
  xleftmargin=18pt,
  xrightmargin=18pt,
  framexleftmargin=6pt,
  framexrightmargin=6pt,
  captionpos=b,
  morekeywords={attacker_control,in_scope,trust_boundary,security_relevant_state,out_of_scope},
  keywordstyle=\color{codeblue}\bfseries,
  morekeywords=[2]{channel,count,profile,transform,state},
  keywordstyle=[2]\color{colChan},
}

\lstdefinestyle{diffstyle}{
  basicstyle=\ttfamily\scriptsize,
  frame=single,
  framerule=0.4pt,
  numbers=left,
  numberstyle=\tiny\color{gray},
  numbersep=5pt,
  xleftmargin=1em,
  breaklines=true,
  columns=fullflexible,
  showstringspaces=false,
  aboveskip=6pt,
  belowskip=3pt,
  captionpos=b,
  escapeinside={(*@}{@*)},
  morecomment=[f][\color{black}\itshape]{//},
  morecomment=[f][\color{codegreen!55!black}]{+},
  morecomment=[f][\color{red!75!black}]{-},
}

\renewcommand{\eqref}[1]{\textup{\hyperref[#1]{(\ref*{#1})}}}

\newcommand{\secref}[1]{\hyperref[#1]{\S\ref*{#1}}}

\begin{document}
\title{\tool: Agentic Vulnerability Detection via Specification Inference}

\author{\IEEEauthorblockN{
 Zhengxiong Luo\IEEEauthorrefmark{2},
Mehtab Zafar\IEEEauthorrefmark{2}, 
Dylan Wolff,
Abhik Roychoudhury }
  \IEEEauthorblockA{
    National University of Singapore
   }
  \IEEEauthorblockA{
   \{{\hypersetup{hidelinks}\href{mailto:luozx@nus.edu.sg}{luozx}},  {\hypersetup{hidelinks}\href{mailto:mehtab@nus.edu.sg}{mehtab}}, {\hypersetup{hidelinks}\href{mailto:abhik@nus.edu.sg}{abhik}}\}@nus.edu.sg, \ {\hypersetup{hidelinks}\href{mailto:wolffd@comp.nus.edu.sg}{wolffdy0}}@gmail.com
   }
  \thanks{\IEEEauthorrefmark{2}Both authors contributed equally.}
 }

\maketitle

\begin{abstract}
The advent of agentic vulnerability detection is already becoming a watershed moment for software security.
Audits conducted entirely by autonomous LLM agents are uncovering critical vulnerabilities in fundamental software that forms the basis of digital society.
Many of these vulnerabilities have remained masked for years and are being uncovered only now with the help of AI agents.
Yet the \emph{reasoning} behind these discoveries remains alarmingly opaque and unvalidated.
What assumptions did the agent make about a function's inputs when it deemed that function to be secure?
Failures in reasoning and incorrect assumptions can lead to missed vulnerabilities and reduce trust in agentic analysis.

In this work, we propose a novel security-specification-first paradigm that (1) exposes the agent's tacit assumptions explicitly as security specifications and (2) continuously refines those specifications via runtime falsification.
We realize our approach in \tool, a novel harness for \emph{agentic vulnerability detection}.
Given a codebase, \tool
analyzes each component of the system for vulnerable code.
When it deems a component to be secure, it commits the local invariants behind that judgment as in-source assertions.
In parallel, \tool leverages a guided fuzzer to attempt to falsify those assumptions.
When the fuzzer triggers an assertion, this either reveals a genuine vulnerability or a flawed specification to refine.
In both cases, this process grounds the agent's understanding, aligning its view of code intent with how the code actually behaves. On real-world subjects, we find that \tool effectively leverages security specifications to detect more vulnerabilities than other state-of-the-art agents.
Additionally, \tool found \bugnum \emph{new} vulnerabilities in key open-source projects, \fixedbugnum of which have already been fixed or confirmed by developers. Compared to curated specialized models like Claude Mythos, our approach presents an effective agentic vulnerability detection approach that can be built on top of widely available LLMs like Sonnet and DeepSeek. 

\end{abstract}


\section{Introduction}

While experts have long decried the inadequacy of security measures in the software industry, recent events have made these warnings impossible to ignore.
Agentic vulnerability detection systems, such as Claude Mythos~\cite{mythos}, are finding more vulnerabilities than ever before~\cite{grinstead_holler_braun_2026}.
Meanwhile, many high-impact vulnerabilities being discovered, such as the Copy-Fail bug in the Linux kernel, have lain dormant for years in high-profile, open-source software.\footnote{\url{https://nvd.nist.gov/vuln/detail/CVE-2026-31431}}~%
As such, it is critical that we understand and harness autonomous security agents to reduce the dangerous security debt accrued in our collective software infrastructure.

Security audits build far more knowledge than the bug reports they deliver.
To find a vulnerability, an analyst must understand the assumptions of the codebase, typically implicit and scattered across disparate contexts, and spot where they might be violated.
Yet the CVE reports that result from this analysis capture only \emph{one instance of a vulnerability}, not this hard-won understanding and analysis.
This loss is consequential: as the discovery of the Copy-Fail bug, which quickly led to the similar Dirty-Frag vulnerability\footnote{\url{https://nvd.nist.gov/vuln/detail/CVE-2026-43284}}, shows, patterns of invalidated assumptions often lead to repeated vulnerabilities.
By focusing only on bug reports, analysts (autonomous agents or otherwise) fail to generalize this knowledge to similar code elsewhere or persist it for future audits, leading to regressions, incomplete fixes, and duplicated analysis.

More broadly, the knowledge built during an audit represents \emph{security specifications} for that project.
While incomplete relative to functional correctness specifications, security specifications in the form of local invariants---predicates that should hold at strategic program points---can encode the critical conditions for maintaining the system's global security.
Unfortunately, in practice, most software comes \emph{without} explicit security specifications, so analysts must \emph{infer} them.

Thus, an ideal agentic analyst must also be able to infer security specifications effectively, not only to find today's vulnerabilities but also to lay the foundation for a secure codebase in the future.
Yet how can we obtain security specifications \emph{autonomously} from \emph{existing} codebases?
How do we gain \emph{confidence} in these specifications?
How do we effectively and automatically \emph{leverage} specifications to aid in finding new bugs?

Large Language Models (LLMs), and in particular agentic systems, have proven effective at inferring functional correctness specifications~\cite{specrover, ma2025specgen}.
It thus stands to reason that they can also infer \emph{security specifications}: indeed, Claude Mythos and other agentic systems~\cite{mythos, bigsleep} are already finding many new vulnerabilities in open-source software, implying some capability for security specification inference.
However, existing agentic analysts infer these specifications only \emph{implicitly}, leaving them both (1)~\textbf{opaque}: the reasoning unfolds in a semantically ambiguous natural-language trajectory that is difficult to understand, let alone reuse in later audits;
and (2)~\textbf{fallible yet largely unvalidated}: an agent's reasoning, like a human expert's, is intuitive rather than systematic, following a few plausible paths and prone to missing corner cases.
Worse still, agents possess limited capability to validate the specifications they infer: even when coupled with dynamic execution~\cite{claudecode}, an agent validates only the suspected \emph{bugs} these specifications lead to, not the \emph{specifications} themselves.
These limitations make a clean audit hard to trust: when an agent reports no bugs, one cannot tell whether the code withstood scrutiny or the implicit specifications behind that verdict were simply incorrect.

\textbf{Our Approach.}
We address these limitations with a security specification-first approach, in which the agent's tacit understanding of the program is made explicit so that it can be continually falsified at runtime and ultimately leveraged to find new bugs.
Our agentic vulnerability detection system, \tool%
, combines the intuitive reasoning of an agent with the more comprehensive exploration of a fuzzing engine, using explicit security specifications to bridge the gap between these two paradigms and provide a comprehensible and reusable artifact from each audit beyond individual vulnerability reports.

Guided by a threat model indicating the system's security boundaries and objectives, \tool begins its audit by analyzing each component, either flagging a vulnerability directly or, when it deems the code secure, determining local invariants at program points in support of that judgment.
\tool durably commits these invariants as executable security specifications to the project's source repository in the form of assertions.
Each local invariant encodes assumptions the agent believes must hold for the system to remain secure and serves a dual role: a \emph{reasoning anchor} at which the agent records its semantic understanding of the program, and a \emph{directional landmark} that comprehensive exploration approaches such as fuzzing can steer toward.

These executable security specifications are directly amenable to validation, for which \tool leverages a complementary fuzzer to falsify them, a process less susceptible to omissions or other failures in LLM reasoning itself.
In turn, the invariants make fuzzing more tractable: an invariant typically lies at a shallower semantic layer than the crash it forestalls, giving the fuzzer an accessible target rather than a deep symptom to stumble upon.
When the fuzzer uncovers a violation of an invariant, this results in one of two outcomes: (\romannumeral1) a \emph{security-relevant violation} indicates an actual defect in the program and surfaces a vulnerability, while (\romannumeral2) a \emph{benign violation} indicates that the agent's security specification needs to be refined, exposing a divergence between the agent's assumptions about the program and the program's actual behavior.
The latter case drives the agent to reassess the corresponding code region and iteratively refine its security specifications before they are checked again by the fuzzer.
This closes the loop between agentic semantic analysis and fuzz-driven edge-case exploration, harnessing each paradigm's strengths to cover the other's weaknesses.

We evaluate \tool on two state-of-the-art benchmarks drawn from DARPA's AI Cyber Challenge (AIxCC)~\cite{aixcc} and the OSV database~\cite{osv}.
Across both benchmarks and under two different frontier LLMs, \tool finds \bugfold more bugs than the state-of-the-art agentic bug detection systems \claude~\cite{claudecode} and the AIxCC-winning \atlantis~\cite{atlantiscrs}, 
with the inferred specifications playing a central role in these discoveries.
Beyond the benchmarks, \tool discovered \bugnum new vulnerabilities in widely-used open-source projects, \fixedbugnum of which have already been fixed or confirmed by developers, with several earning bug bounty rewards.
Finally, we showcase that the inferred invariants are valuable, durable artifacts that outlast a single audit, e.g., pinpointing the incomplete fix and bug family behind a four-month series of fixes in the actively developed project \texttt{gpsd}~\cite{gpsd}.

In summary, we make the following contributions:
\begin{itemize}[leftmargin=*]
    \item We formulate a novel security-specification-first paradigm for agentic vulnerability detection, which makes the agent's tacit security judgments explicit, falsifiable, and continually aligned with the program's actual behavior.
    \item We realize this paradigm in \tool: the agent commits its assumptions as in-source invariants, and a guided fuzzer continuously falsifies them to expose flawed assumptions, forming a reason-falsify-refine loop.
    \item We evaluate \tool and demonstrate its substantial improvement in bug finding over state-of-the-art agentic systems, uncovering \bugnum previously unknown vulnerabilities in widely-used projects (\fixedbugnum fixed or confirmed, several bounties awarded).
\end{itemize}

\section{Motivation}
\label{sec:motivation}

To motivate the design of \tool, we examine an example vulnerability from the recent AIxCC challenge~\cite{aixcc} and analyze the limitations of existing work.

\begin{figure*}[t]
  \centering
  \includegraphics[width=\textwidth]{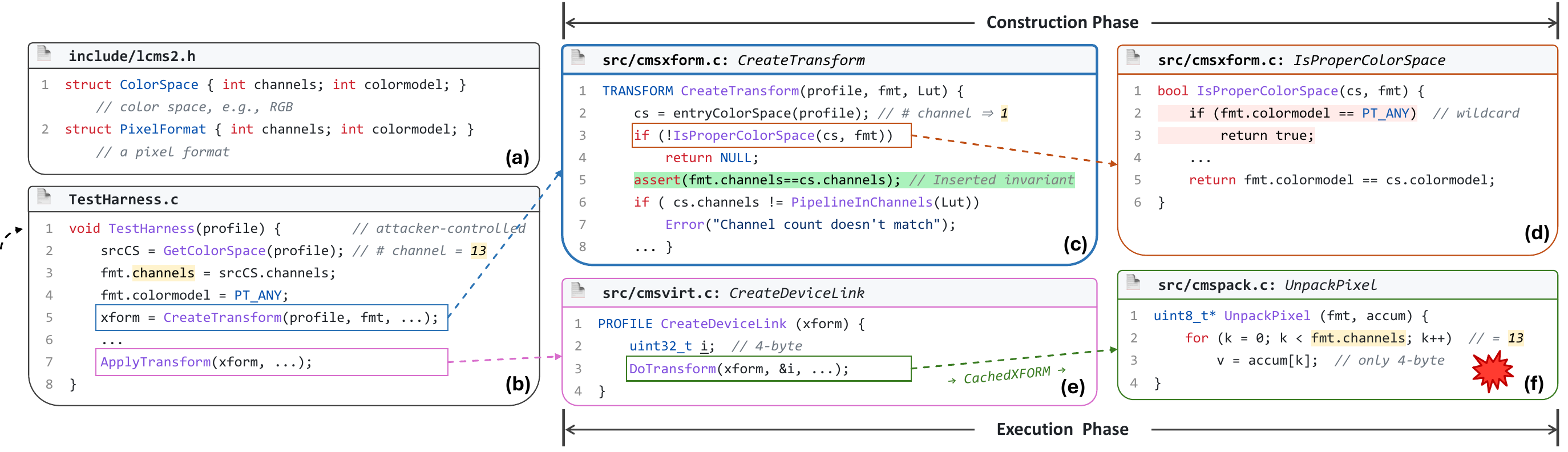}
  \caption{ The simplified code snippets of the bug example in \texttt{Little CMS}. In the \emph{construction phase} (above), 
  an attacker-controlled profile builds a pixel format with 13 channels and the \texttt{PT\_ANY} wildcard, while the library fixes the transform's entry color space to a single-channel space. 
  \texttt{IsProperColorSpace} accepts the format through its \texttt{PT\_ANY} branch without comparing channel counts, so the agent regards the construction as secure; \tool can commit this assumption as the local invariant $\phi$,
  \texttt{fmt.channels} $=$ \texttt{cs.channels}, which the fuzzer may later falsify when the transform is created, at which point no memory error is visible. In the \emph{execution phase} (below), the same input carries the inconsistent channel count through a separate, later call into \texttt{UnpackPixel}, which can read past the 4-byte buffer and trigger an out-of-bounds read.
  Dashed arrows are call-graph edges; the inconsistent channel count (13) is highlighted
  along its path from source to sink.}
  \label{fig:motivating-example}
\end{figure*}

\subsection{Motivating Example}
\label{sec:motivation:example}

Figure~\ref{fig:motivating-example} walks through the challenge in \texttt{Little CMS}, a widely used color-management library.
\texttt{Little CMS} converts \emph{pixels} between \texttt{ColorSpaces} (e.g., RGB, GRAY). A conversion pairs (1) a \texttt{ColorSpace} with (2) a \texttt{PixelFormat} recording how the input bytes are packed (Figure~\ref{fig:motivating-example}a). Since \texttt{PixelFormat} and \texttt{ColorSpace} are two views of the same \emph{pixels}, they must agree on the \emph{channel} count: the count $\Cfmt$ the \texttt{PixelFormat} declares must equal the count $\Ccs$ the \texttt{ColorSpace} implies, i.e., $\Cfmt = \Ccs$. Otherwise, the code that later unpacks a \emph{pixel} would read the wrong bytes.

At a high level, the \texttt{TestHarness}~(Figure~\ref{fig:motivating-example}b) reads an input from a user-supplied
color profile~(lines 1-4) and passes it to the \emph{construction phase} via \texttt{CreateTransform}~(line 5), which checks the
input and assembles a \texttt{Transform}; only once these checks pass does the \emph{execution phase} execute such a \texttt{Transform} by calling \texttt{ApplyTransform}~(line 7).

Figure~\ref{fig:motivating-example}c shows the {construction phase}: it validates the \texttt{PixelFormat} \texttt{fmt} against the \texttt{ColorSpace} \texttt{cs} using the guard \texttt{IsProperColorSpace}~(line 3, Figure~\ref{fig:motivating-example}d). This function compares the \texttt{colormodel} of \texttt{fmt} and \texttt{cs}: a matching \texttt{colormodel} typically implies a matching \emph{channel} count. 
After that, \texttt{CreateTransform} also compares $\Ccs$ against a {different} source---the conversion's internal pipeline, and rejects any mismatch with the error \texttt{``Channel count doesn't match''}~(lines 6-7).
The validation therefore looks sound, and the LLM agent, reading the same code, concludes likewise and regards it as secure.

However, this reasoning is \emph{incorrect} due to a branch within the \texttt{IsProperColorSpace} guard: as shown in lines 2--3 in Figure~\ref{fig:motivating-example}d, for a special \texttt{colormodel} \texttt{PT\_ANY}, the guard returns \texttt{true} directly, since \texttt{PT\_ANY} is a wildcard that matches any \texttt{colormodel}. 
Therefore, for a \texttt{TestHarness} carrying exactly this \texttt{colormodel} \texttt{PT\_ANY}~(Figure~\ref{fig:motivating-example}b, line 4), the intended agreement $\Cfmt = \Ccs$ is never enforced.
In this scenario, a carefully crafted profile input could push the two counts apart: (\romannumeral1) its declared color space sets \texttt{fmt.channels} to $13$~(Figure~\ref{fig:motivating-example}b, lines 2--3), so $\Cfmt = 13$; and (\romannumeral2) other fields of the input profile make \texttt{entryColorSpace}~(Figure~\ref{fig:motivating-example}c, line 2) return a single-channel \texttt{ColorSpace}, so $\Ccs = 1$. 

The divergence between $\Cfmt$ and $\Ccs$ has no immediate ill effects: the mismatched counts are stored as ordinary integers, and nothing in the construction phase signals an error. Whether the divergence manifests later as a visible bug depends on how the stored \texttt{PixelFormat} is used.
In the execution phase,
\texttt{ApplyTransform}~(Figure~\ref{fig:motivating-example}b) routes the inflated \texttt{PixelFormat} onto a path provisioned for unpacking a small pixel (Figure~\ref{fig:motivating-example}e), where the unpack loop~(Figure~\ref{fig:motivating-example}f) iterates the declared $\Cfmt = 13$ times over a 4-byte buffer~(\texttt{uint32\_t i}), causing an out-of-bounds read that AddressSanitizer~\cite{Serebryany2012AddressSanitizerAF} reports.

Identifying this vulnerability is extremely difficult: the causal chain from the bypassed validation check to the crash-site spans two phases of execution, multiple call
layers, and separate source files.

\subsection{Limitations of Existing Work}
\label{sec:motivation:limitations}
Whether an analysis tool can detect a bug like the one in Figure~\ref{fig:motivating-example} comes down to how faithfully its understanding of the program matches the program's actual semantics.
For a target program $P$, let $I$ be the in-scope
inputs an attacker can supply, and $\beh{P}{I}$ be the executions $P$ admits on them. A real vulnerability is an execution that reaches an error state, collected in $\viol$; the ground-truth bugs are thus:
\[
    \beh{P}{I} \,\cap\, \viol
\]

\begin{figure}[H]
    \centering
    \includegraphics[width=.9\linewidth]{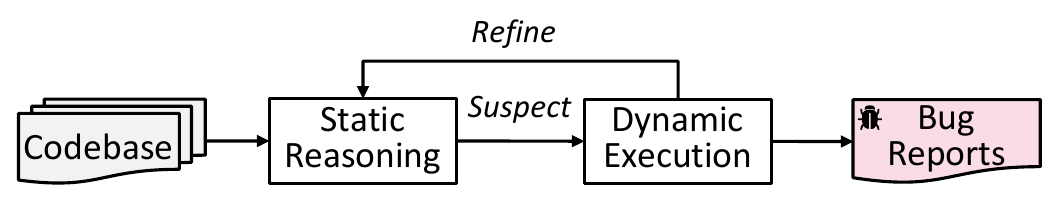}
    \caption{Existing paradigm of agentic bug detection: the agent reasons and flags suspects for dynamic validation. However, the underlying \emph{implicit} security specification itself remains \emph{largely unvalidated}.}
    \label{fig:existing_CRS}
\end{figure}

\textbf{Agentic Approach.} An LLM-based analyst reasons not over the real behaviors $\beh{P}{I}$ but over its own \emph{view} of program $P$, written $\behm{P}{I}$: the executions its reasoning reconstructs, over the input space $I$.
The ideal case is $\behm{P}{I} = \beh{P}{I}$, but this is hard to achieve especially for complex programs with intricate dynamic behaviors.

The prevailing approach to closing this gap is to selectively ground the agent via simple tools such as code execution~\cite{claudecode}.
As Figure~\ref{fig:existing_CRS} shows, the agent first reasons statically and flags each \emph{suspect}, an input $\isus \in I$ it believes can reach an error state, then validates it by dynamic execution.
This workflow leads to a dual view for the agent:
\begin{empheq}[left={\behm{P}{I}\;\xrightarrow{\substack{\emph{agentic}\\\emph{capability}}}\;\empheqlbrace}]{align}
    \beh{P}{\isus}  &\quad \text{for \emph{suspect} } \isus \in I, \label{eq:ground-susp}\\
    \behm{P}{i} &\quad \text{for \emph{non-suspect} } i \in I. \label{eq:ground-nonsusp}
\end{empheq}
It reports a vulnerability whenever a suspect $\isus$'s run genuinely reaches an error state, i.e., $\beh{P}{\isus} \;\cap\; \viol \neq \emptyset$. Meanwhile, the agent's view $\behm{P}$ is continuously refined by what concrete executions reveal during this process.

Although this agentic approach is powerful and has uncovered many vulnerabilities, it still faces major limitations.
In case~\eqref{eq:ground-nonsusp}, the agent deems a site secure and never raises it as a suspect, so the dynamic stage is never invoked to test it.
Put differently, the agent's view $\behm{P}$ acts as an \emph{implicit security specification}: dynamic execution validates only the suspected bugs flagged by this specification, i.e., case~\eqref{eq:ground-susp}, not the specification itself, which harbors the seemingly secure, non-suspect verdicts, i.e., case~\eqref{eq:ground-nonsusp}.
In our example (Figure~\ref{fig:motivating-example}), the agent follows the guard \texttt{IsProperColorSpace} along plausible paths and (incorrectly) infers that this guard \emph{implies} a matching channel count, deeming \texttt{CreateTransform} secure.
Fundamentally, the security of a program point is not a property of any single path, but must hold over \emph{all} paths that reach it, under \emph{diverse} inputs. An agent may reason strongly along one or a few individual paths, yet a sound security judgment must account for that entire combinatorial space, which is far harder.

\textbf{Brute-Force Approach.} A brute-force approach, such as fuzzing~\cite{miller1990empirical}, instead randomly samples $\beh{P}{I}$ directly and reports only bugs it can trigger in $\beh{P}{I} \cap \viol$. 
By virtue of this randomized sampling process, fuzzing often surfaces counter-intuitive inputs that developers and security auditors overlook. 
However, lacking semantic reasoning, blind randomization struggles to reach deep program states. 
In our example (Figure~\ref{fig:motivating-example}), producing a proof-of-vulnerability means satisfying several narrow constraints simultaneously: building a semi-valid \texttt{profile} that passes validation, declaring more channels than the data carries, and carrying that \texttt{profile} through the execution phase into \texttt{UnpackPixel} to trigger the out-of-bounds read.
This semantic blindness also extends to the channel-count mismatch when it occurs at construction (Figure~\ref{fig:motivating-example}c).
A fuzzer does not know that this mismatch constitutes an interesting state, and thus even grey-box fuzzers~\cite{libfuzzer} cannot bias sampling that state to alleviate the difficulty of satisfying all constraints necessary to manifest the bug.

\section{Our approach}

Our \textbf{goal} is to make the LLM agent's view of the program \emph{more} faithful to reality, i.e., in the notation of \secref{sec:motivation:limitations}:
\begin{equation}
    \behm{P}{I} \; \leadsto \; \beh{P}{I},
    \label{eq:goal}
\end{equation}

To fill this gap, the agent that pronounces a site secure must itself be audited: we must check whether its security judgments actually hold, i.e., case~\eqref{eq:ground-nonsusp}, the non-suspect region where false negatives hide.
This task is challenging at two levels: \emph{understanding} the reasoning behind the judgment and \emph{validating} the understanding. 

\textbf{Challenge 1}:~\emph{Understanding} an agent's judgment is difficult because the reasoning behind it is \emph{implicit}, reached through a semantically ambiguous natural-language reasoning trajectory. 
Moreover, this reasoning is couched in the program's own intent~(e.g., that a color space and a pixel format must agree on channel count, as shown in \secref{sec:motivation:example}) and states~(e.g., \texttt{fmt.channels} and \texttt{cs.channels}), so interpreting it correctly requires grasping these program-specific concepts, which any \emph{separate} formalism (e.g., a bespoke specification language) would have to re-encode before the claim could even be stated, a step that is both prone to information loss and inefficient.

\textbf{Challenge 2}:~\emph{Validating} is often equally difficult, as it typically involves reaching the precise conditions under which an incorrect judgment will break in a real execution.
Neither of the two natural validators suffices by itself:
(a)~An LLM-based validator shares the same blind spots that produced the judgment in the first place, and tends to overlook the same corner cases a second time.
(b)~A brute-force approach, such as a fuzzer, is unaware of program semantics, rarely reaching deep states and making it difficult to distinguish between semantically relevant and irrelevant states among the many reachable states.

\textbf{Basic Idea.}
\tool pursues the goal~\eqref{eq:goal} with a security-specification-first approach to these two challenges.

(\romannumeral1) For \emph{understanding}, rather than recovering the agent's implicit, program-specific reasoning trajectory after the fact, \tool lets the agent write its understanding down \emph{explicitly}: whenever the agent deems a site secure, it states the assumption behind that verdict as an \emph{invariant} $\phi$. $\phi$ is a falsifiable assertion placed at the site in the program's own source (e.g., Figure~\ref{fig:motivating-example}c, line~5).  
By encoding assumptions as local invariants, \tool ensures that the assumption becomes an obligation over \emph{any} path that reaches the site, instead of a statement about one particular path (\secref{sec:motivation:limitations}).
Writing $\phi$ as an assertion in the program's source also spares \tool from re-encoding each project's concepts into a separate formalism.
However, the most important advantage of an \emph{executable specification} is for \emph{validation}.

(\romannumeral2) For \emph{validation}, \tool grounds $\phi$ via executing the program under a large number of automatically generated test cases.
Rather than using the agent itself to generate test cases, which would be susceptible to the same blind spots that might lead to invalid invariants in the first place, \tool delegates test generation to a grey-box fuzzer.
\tool makes $\phi$ the fuzzer's target to \emph{falsify}, searching for a witness input $i_{\neg\phi}$ whose execution trace satisfies $\neg\phi$. Because $\phi$ pins down only \emph{what} must hold and not \emph{how} to reach the site, the fuzzer is free to violate it by \emph{any} path. Steering the fuzzer by $\phi$ rather than by blind coverage is more effective in two ways. First, the search aims squarely at falsifying a given property $\phi$ instead of widening coverage indiscriminately. Second, $\phi$ typically sits at a higher semantic layer than the eventual crash, so the fuzzer pursues a shallow, semantic goal~(reaching the inconsistent state that breaks $\phi$ in Figure~\ref{fig:motivating-example}c) rather than a deep symptom to stumble upon~(the out-of-bounds read in Figure~\ref{fig:motivating-example}f). A resulting violation of $\phi$ surfaces the divergence between the agent's model $\behm{P}{I}$ and the program's actual behavior $\beh{P}{I}$ and serves as a \emph{lead} that guides the agent to refine its understanding of the code. In our example in \secref{sec:motivation:example}, the agent's belief that $\texttt{fmt.channels} = \texttt{cs.channels}$~(Figure~\ref{fig:motivating-example}c) can become the invariant $\phi$, which a fuzzer-generated 13-versus-1 \texttt{profile} input $i_{\neg\phi}$ later falsifies, surfacing the \texttt{PT\_ANY} bypass missed by purely static reasoning.

\section{System Design}

\begin{figure}[t]
    \centering
    \includegraphics[width=\linewidth]{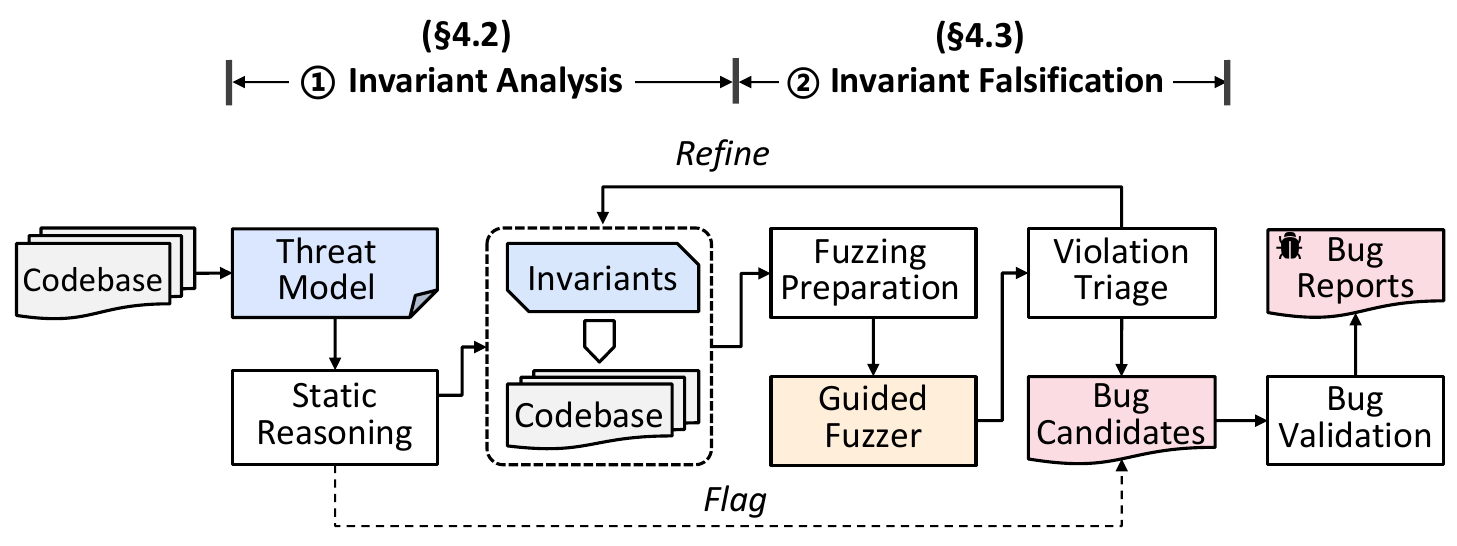}
    \caption{Overview of \tool, which turns the agent's \emph{implicit} reasoning into \emph{explicit}, \emph{falsifiable} security specifications.
    \textbf{\ding{172}} Given a codebase, \tool distills a \hlbox{workflowblue}{threat model} capturing the project's high-level security intent and context. Guided by it, \tool inspects the code and either \emph{flags} \hlbox{workflowpink}{bug candidates} directly or, deeming a site secure, \emph{commits} the supporting \hlbox{workflowblue}{invariants} as in-source assertions.
    \textbf{\ding{173}} These beliefs the agent cannot guarantee become checks a \hlbox{workfloworange}{\texttt{Guided Fuzzer}} can run and falsify, and each violation, once triaged, yields either \hlbox{workflowpink}{bug candidates} or flawed assumptions to {refine}.
    Finally, \tool validates each candidate before reporting.
    }
    \label{fig:overview}
\end{figure}

Figure~\ref{fig:overview} shows \tool{}'s workflow.
Given a codebase $P$, \tool first distills a threat model from the code, its documentation, and its build context.
This threat model captures the project's security boundary (e.g., the attack surface and trust boundaries) and the security-relevant properties that code within this boundary must uphold, serving as the global context about the project for the subsequent audit.

Guided by the threat model, \tool enters the \textbf{Invariant Analysis} stage~(\ding{172}), where \texttt{Static Reasoning} inspects the code, asking what each site should guarantee under this security boundary.
When a site appears insecure, \tool flags it directly as a \emph{bug candidate} (along the dashed ``\emph{Flag}'' path), as a conventional agentic audit would.
When a site is deemed secure, \tool instead records the assumptions behind this verdict as invariants, each committed into the source as an assertion, yielding the instrumented program $P'$.

In \textbf{Invariant Falsification}~(\ding{173}), \tool leverages grey-box fuzzing to run $P'$, seeking to falsify any committed invariant $\phi$.
\texttt{Fuzzing Preparation} builds $P'$ into a fuzzing target whose feedback rewards progress toward a violation $\neg\phi$.
Guided by this feedback, the \texttt{Guided Fuzzer} searches for a violating input.
\texttt{Violation Triage} then inspects each violation.
A violation that reflects a genuine security issue becomes a bug candidate.
Such a violation is not always the vulnerability itself: it is often an early inconsistency that leads the agent to a downstream defect, as in our motivating example (\secref{sec:motivation:example}).
A benign divergence, by contrast, reveals a flaw in the agent's understanding of the code rather than a bug. \tool then updates that understanding, refining this invariant as necessary.
This reason-falsify-refine loop progressively aligns the agent's view of the program with the program's actual behavior.

Finally, every bug candidate is validated before being reported.
The audit's final artifacts are the threat model, the surviving invariants, and the validated bug reports.

\subsection{Threat Model Construction}
\tool begins by distilling the global context for the audit, including the threat model.
Reading the codebase, its documentation, and its build configuration, \tool reasons at a high level, aiming to capture the security properties the system is meant to preserve rather than the details of any single function.
It records the threat model as a structured document.
We show an example snippet of the threat model generated for the \texttt{Little CMS} running example in Listing~\ref{lst:lcms-threat-model}.
Notably, we can see that in this case, \tool identifies that the pixel format and channel count are \emph{important} variables within the global context, helping it later deduce the critical \emph{local} invariant that 
$\Cfmt = \Ccs$.

\begin{lstlisting}[
  style=tmstyle,
  caption={Threat-model fragment for the motivating example.},
  label={lst:lcms-threat-model}
]
attacker_control: input ICC profile bytes
in_scope: transform construction, pixel unpacking
trust_boundary: profile fields -> transform state
security_relevant_state:
  - pixel format channel count
  - color space channel count
out_of_scope:
  ...
\end{lstlisting}

A threat model contains many different fields, each with different purposes throughout the audit.
For example, the \texttt{security\_relevant\_state} field lists variables that relate to possible global security properties.
The \texttt{attacker\_control} and \texttt{in\_scope} fields tell \tool which inputs and execution paths matter, \texttt{trust\_boundary} marks where invariants belong, and \texttt{out\_of\_scope} bounds which failures bug validation may report.
Together the threat model is used throughout the overall workflow, guiding the static reasoner towards in-scope and relevant source locations, giving relevant state variables from which to build invariants during invariant analysis, and giving context to the validation and triage steps for what constitutes a valid report for a given project.

\subsection{Invariant Analysis}
\label{sec:design:analysis}

\begin{algorithm}[t]
\DontPrintSemicolon
\small
\SetCommentSty{algcommentstyle}
\SetKwInOut{Input}{Input}
\SetKwInOut{Output}{Output}
\SetKwFunction{Falsify}{Falsify}
\Input{$P$ -- program under audit \\ $T$ -- threat model}
\Output{$B$ -- bug candidates\\ $\Phi$ -- surviving invariants}
$\Phi \gets \emptyset$,\quad $B \gets \emptyset$,\quad $C \gets \emptyset$\;
\Repeat{audit budget is exhausted}{
  \tcc{Static Reasoning: reason about a site against $T$}
  $(h, s) \gets$ \textsc{GenerateHypothesis}($P$, $T$, $\Phi$)\label{alg:ana:hypo}\;
  $(b, \phi) \gets$ \textsc{EvaluateHypothesis}($P$, $T$, $h$, $s$)\label{alg:ana:reason}\;
  \uIf{$b \neq \bot$}{$B \gets B \cup \{b\}$\label{alg:ana:flag} \tcp*{insecure site: flag}}
  \uElseIf{$\phi \neq \bot$}{
    \Repeat{$ok$}{
      $P' \gets$ \textsc{Instrument}($P$, $\Phi \cup \{\phi\}$)\label{alg:ana:instrument}\;
      $(ok, e) \gets$ \textsc{Check}($P'$, $\phi$)\label{alg:ana:check}\;
      \lIf{$\neg ok$}{$\phi \gets$ \textsc{RepairAssertion}($\phi$, $e$)}
    }
    $(b, fb, C) \gets$ \Falsify{$P$, $P'$, $\phi$, $C$}\label{alg:ana:falsify} \;
    \uIf{$b \neq \bot$}{$B \gets B \cup \{b\}$ \tcp*{a real bug}}
    \uElseIf{$fb \neq \bot$}{
   $\phi \gets$ \textsc{RefineInvariant}($T$, $\phi$, $fb$)\label{alg:ana:refine} \;
    }
    \Else{$\Phi \gets \Phi \cup \{\phi\}$ \tcp*{$\phi$ survives}\label{alg:ana:hold}}
  }
}
\Return $B$, $\Phi$\;
\caption{Iterative Invariant Analysis}
\label{alg:analysis}
\end{algorithm}

Algorithm~\ref{alg:analysis} details \tool's \emph{iterative} invariant analysis, whose LLM-driven steps are  \textsc{GenerateHypothesis}, \textsc{EvaluateHypothesis}, and \textsc{RefineInvariant}.
Given the codebase $P$ and the threat model $T$, \tool repeats a reason-falsify-refine cycle, flagging bug candidates $B$ and accumulating the invariants $\Phi$ that survive to later iterations.

\textbf{Static Reasoning.}
In \texttt{Static Reasoning}, \tool forms hypotheses regarding the source code against the threat model (\textsc{GenerateHypothesis}, line~\ref{alg:ana:hypo}).
Each hypothesis $h$ indicates a developer assumption about a location $s$ (e.g., a function) that it proposes can be broken.
These hypotheses are grounded on one side by the source code $s$ and on the other by an attacker capability declared by the threat model.
After forming $h$, \tool first evaluates it statically, using iterative LLM-based reasoning to either flag a hypothesis as a true candidate or as unreachable within the context of the project (\textsc{EvaluateHypothesis}, line~\ref{alg:ana:reason}).

For example, in \texttt{Little CMS}, the threat model states that the attacker controls the ICC-profile bytes that flow into transform construction (Listing~\ref{lst:lcms-threat-model}).
Based on this information, \tool might hypothesize when analyzing \texttt{CreateTransform} (Figure~\ref{fig:motivating-example}c) that a named-color profile passes the color-space guard while storing a pixel format whose channel count differs from the transform's color space.
The distinction between local behavior and intended property here is critical;
seeing the call to \texttt{IsProperColorSpace} in isolation might lead the model to believe that \emph{the color models match}.
But as we saw in \secref{sec:motivation:example}, this guard statement is \emph{insufficient} to uphold the intent recorded in the threat model. 

If the hypothesis is deemed plausible by this evaluation, \tool \emph{does not} directly report it.
This is because hallucinations and other errors in LLM-reasoning at this stage could lead to false positive bug reports.
Instead, \tool emits a bug candidate $b$ (line~\ref{alg:ana:flag}) to be \emph{validated} by a concrete execution (\secref{sec:design:val}).
If \tool instead deems the hypothesis to be invalid, it commits a falsifiable invariant $\phi$ that records the assumption on which that verdict rests, described below.

\textbf{Committing Invariants.}
Each committed $\phi$ is tied to the threat model through a trust boundary or attack surface.
Additionally, the assertions are not arbitrary functional properties but the conditions the threat model identifies as \emph{security-relevant}.
\tool inserts each $\phi$ into the project source as a tagged assertion and rebuilds the project as $P'$ without any further change to its build configuration (line~\ref{alg:ana:instrument}).
The assertion takes whatever form is native to the project's source language: sanitizer-style traps for C and C++, standard assertions for Go, Jazzer\cite{jazzer}-compatible feedback channels for the JVM, and so on.
The only requirement is that an invariant violation at runtime is structurally distinguishable from other program failures, so it can be identified and triaged later.
Each invariant $\phi$ therefore carries a unique identifier.
After instrumentation, \tool checks that $P'$ builds successfully and that the inserted assertion is syntactically valid (line~\ref{alg:ana:check}), repairing the assertion and rebuilding until it is valid.
This iterative build-time repair only fixes a malformed assertion, and is distinct from the semantic refinement in \secref{sec:design:falsify}.

Notably, these invariants $\Phi$ can generalize far beyond an individual bug report, which we examine further in \secref{sec:eval:rq4}.
For example, due to its placement in the source code, the inferred \texttt{Little CMS} invariant guards other, similar conversions present in the library but not explored via \texttt{TestHarness.c}.
Indeed, the same predicate can be re-checked on the symmetric output-format path, on other entry points, and on later revisions of the library, flagging code that re-introduces the wildcard bypass.

\subsection{Invariant Falsification}
\label{sec:design:falsify}

\begin{algorithm}[t]
\DontPrintSemicolon
\small
\SetCommentSty{algcommentstyle}
\SetKwProg{myproc}{Procedure}{}{}
\SetKwFunction{Falsify}{Falsify}
\myproc{\Falsify{$P$, $P'$, $\phi$, $C$}}{
  \tcc{$P$: original program; $P'$: instrumented program; $\phi$: target invariant; $C$: seed corpus}
  \tcc{Fuzzing Preparation: build target, test harness, extend corpus}
  $(\mathcal{H}, C) \gets$ \textsc{FuzzingPreparation}($P'$, $C$)\label{alg:fal:prep} \;
  \tcc{Specification Guided Fuzzer: seek a witness that violates $\phi$}
  $(i_{\neg\phi}, C) \gets$ \textsc{SpecGuidedFuzz}($P'$, $\mathcal{H}$, $\phi$, $C$)\label{alg:fal:fuzz}\;
  \lIf{$i_{\neg\phi} = \bot$}{\Return $(\bot, \bot, C)$\tcp*[f]{$\phi$ holds}}
  \tcc{Violation Triage: a real bug, or $\phi$ too strong?}
  $(b, fb) \gets$ \textsc{Triage}($i_{\neg\phi}$, $\phi$, $P$, $P'$)\label{alg:fal:triage}\;
  \Return $(b, fb, C)$\;
}
\caption{Invariant Falsification}
\label{alg:falsification}
\end{algorithm}

\textsc{Falsify} (Algorithm~\ref{alg:falsification}) is the \emph{asynchronous} subroutine the main loop invokes on each committed invariant (Algorithm~\ref{alg:analysis}, line~\ref{alg:ana:falsify}).
It takes the original program $P$, the instrumented program $P'$, the target invariant $\phi$, and the running seed corpus $C$, and returns whether the fuzzer found a real defect, a too-strong invariant, or no violation at all with a justification feedback $fb$.
Recall from \secref{sec:motivation:limitations} that, for an input $i \in I$, $\beh{P}{i}$ denotes the executions $P$ admits on $i$.
If $\phi$ holds, the corresponding assertion instrumentation should not change the program's behavior on any input, i.e., $\beh{P}{i} = \beh{P'}{i}$.
The \texttt{Guided Fuzzer} aims to validate this by searching $P'$ for a witness $i_{\neg\phi}$ that violates $\phi$, i.e., $\beh{P}{i_{\neg\phi}} \neq \beh{P'}{i_{\neg\phi}}$.

\textbf{Falsifying and Refining.}
Each committed invariant is handed straight to the \emph{asynchronous} \textsc{Falsify} subroutine (line~\ref{alg:ana:falsify}), which fuzzes $P'$ for an input that breaks $\phi$ and returns one of three outcomes.
If no input breaks $\phi$, the invariant holds and joins the surviving set $\Phi$ (line~\ref{alg:ana:hold}).
If the breaking input is a genuine defect, it becomes a bug candidate in $B$.
If $\phi$ was merely too strong, i.e., the behavior that violates it is in fact secure, the gap lies in \tool's understanding rather than in the code, so \tool refines the invariant $\phi$ (\textsc{RefineInvariant}, line~\ref{alg:ana:refine}).
This reason-falsify-refine loop runs until the audit budget is exhausted.

\textbf{Grey-Box Fuzzing.}
\tool achieves falsification via \emph{fuzzing}~\cite{miller1990empirical}.
Fuzzing is a randomized search over the input domain of a program (the search space) with the goal of finding inputs that trigger vulnerabilities.
The search process for \emph{grey-box} fuzzers is adaptive, biasing the search towards interesting inputs using a feedback function.
The feedback for a given input involves an assessment of whether novel program behavior was exposed by that input's execution.
Typically, novelty is determined by a measure of code coverage, checking whether new program locations instrumented at compile time were executed at runtime. 

\textbf{Fuzzing Preparation.}
Before any campaign within the broader audit, \tool builds the instrumented source code $P'$ into a runnable grey-box fuzzing target (line~\ref{alg:fal:prep}).
As part of the build process, \tool uses an existing fuzz test harness $\mathcal{H}$ when one is available, refining it as needed, or constructs a new test harness otherwise.
It also wires each committed invariant into the fuzzer's feedback mechanism such that \emph{approaching} an invariant results in gradual, observable progress for the fuzzer.
\tool persists a corpus of interesting fuzzing inputs $C$ across calls to \textsc{Falsify}, so exploration is cumulative as invariants are committed and refined.

As part of the transformation from $P$ to $P'$, $P'$ aborts whenever a committed invariant is \emph{first} violated by a fuzzer input, triggering triage (discussed at the end of this subsection).
Subsequent triggers of the invariant do not immediately abort, allowing the fuzzer to progress and find crashes beyond the initial violating state.
Moreover, each invariant exports a feedback channel in $P'$ for when the invariant itself has been tripped, whether it has been reached, and how close the invariant is to tripping (via a bucketed distance function when available, described in \secref{sec:impl}).
As a result, rather than being purely guided by code coverage, as is typical for grey-box fuzzers~\cite{libfuzzer}, the engine treats progress toward an inconsistent state as interesting behavior and steers toward it.
These invariants also represent an easier-to-reach goal than directly finding a crash.

\textbf{Specification Guided Fuzzing.}
The \texttt{Specification Guided Fuzzer} takes the instrumented target $P'$ and the invariant $\phi$, and searches for a witness $i_{\neg\phi}$ that violates it (line~\ref{alg:fal:fuzz}).
It exercises $P'$ with an off-the-shelf engine such as libFuzzer~\cite{libfuzzer} or Jazzer~\cite{jazzer}.
Rather than adapting the fuzzer \emph{implementation}, \tool instead adapts the target's failure surface and feedback.
This design choice allows for \tool to adopt new, more advanced fuzzing algorithms and implementations in the future for more effective falsification.
Both sanitizer crashes and invariant violations found by the fuzzer
are handed to the triage component.

\textbf{Violation Triage.}
\texttt{Violation Triage} takes a witness $i_{\neg\phi}$ and its invariant tag and uses LLM-based reasoning to decide which of two interpretations applies (line~\ref{alg:fal:triage}).
Either $i_{\neg\phi}$ exposes a real bug in $P$, or $\phi$ was too strong and the violation is benign.
To decide, \tool reproduces the violation and inspects the program state at the assertion site.
When the underlying behavior is genuinely insecure, it returns a bug candidate $b$, carrying the violating input and the invariant tag.
Otherwise it returns feedback $fb$ that explains why $\phi$ was too strong, which the main loop uses to refine the invariant (\textsc{RefineInvariant}, line~\ref{alg:ana:refine}).

\subsection{Bug Validation}
\label{sec:design:val}

The final stage turns the bug candidates $B$ into reported vulnerabilities $R$ by validating them.

\textbf{Proof-of-Vulnerability Construction.}
A candidate surfaced by the fuzzer already carries a witness input, but a statically flagged candidate does not.
For the latter, \tool first constructs a Proof of Vulnerability (PoV): an input that drives the defect from a program-boundary entry point rather than by invoking an internal function directly.
Anchoring the PoV at the boundary establishes that the defect is reachable the way a real attacker would reach it, not through an interface no adversary controls.
If \tool cannot produce a boundary PoV for a given bug candidate, it does not report the bug.

\textbf{Confirming the Candidates.}
With a reproducing input in hand, \tool decides whether the candidate is a genuine vulnerability.
\tool confirms the bug \emph{only} if it deems it to be reachable under the attacker capabilities as declared by the threat model.
Many reports, even those with a PoV, may fall outside the attacker's capabilities due to following a code path outside the in-scope attack surface.
In these cases, \tool does not surface them as vulnerabilities.

\section{Implementation}
\label{sec:impl}

\tool is implemented on top of \texttt{pi v0.77.0}, an AI agent toolkit~\cite{pi}, in 9.1K lines of TypeScript for the runtime and subagent orchestration.
The subagents have access to standard coding tools, including \texttt{read\_file}, \texttt{edit\_file}, Language Server Protocol queries, \texttt{bash}, and \texttt{grep}, as well as workflow-specific tools such as \texttt{spawn\_subagent} and \texttt{steer\_subagent} for controlling subagents and \texttt{complete\_stage} for reporting stage completion.
We run all \tool agents and subagents with the model's thinking disabled and temperature set to $0$.

\myparagraph{Fuzzing and Instrumentation}
We use off-the-shelf grey-box fuzzing tools for each language in our evaluation: libFuzzer~\cite{libfuzzer} for C/C++ programs, Jazzer~\cite{jazzer} for the JVM, and the native fuzzing drivers for Go and Rust. 
\tool adds invariants as \emph{source annotations} in the target language.
For a subset of binary-operator invariants with numeric arguments, \tool also emits fuzzer-specific annotations, when available, to indicate how close the current execution is to violating the invariant.
For example, if an invariant consists of \texttt{len $\leq$ max}, the distance is the absolute value of the difference between \texttt{len} and \texttt{max}.
In the C/C++ backend, each invariant uses 16 libFuzzer extra-counter slots: one for reachability, one for tripping, and 14 logarithmically sized range buckets for numeric distances.
In the JVM backend, \tool uses Jazzer's native \texttt{exploreState} for reachability and \texttt{minimize} for the distance-to-violation value~\cite{jazzer}. In both cases, these signals \emph{do not} modify the fuzzing algorithm; they only make the inferred safety boundary \emph{visible} to the existing coverage-guided search. For other backends such as Go and Rust, \tool currently does not support distance-guided feedback, though it can be easily extended to do so.


\section{Evaluation}
\label{sec:eval}

We evaluate \tool to answer the following research questions:
\begin{itemize}[leftmargin=2.2em]
    \item[\textbf{RQ1}] \textbf{(Effectiveness)} How effective is \tool compared to other agentic baselines in bug discovery?
    \item[\textbf{RQ2}] \textbf{(Component Contribution)} How does each component of \tool contribute to its effectiveness?
    \item[\textbf{RQ3}] \textbf{(Security Impact)} Is \tool effective in exposing unknown vulnerabilities in real-world software?
    \item[\textbf{RQ4}] \textbf{(Usefulness of Artifacts)} How useful are the inferred security specifications across the software's lifecycle?
\end{itemize}

\subsection{Evaluation Setup}
\label{sec:eval_setup}

\textbf{Benchmarks.}
We evaluate \tool on two complementary benchmarks of known vulnerabilities in real-world software.
AIxCC is a \emph{controlled} benchmark of curated vulnerabilities seeded into mature open-source projects, whereas OSV is an \emph{in-the-wild} benchmark of vulnerabilities independently discovered in deployed software.
In both, every vulnerability comes with a reproducible PoV, so that each finding can be verified against a known ground truth.

\begin{enumerate}[leftmargin=*]
    \item \textbf{The AIxCC benchmark}~(Table~\ref{tab:aixcc}) is drawn from DARPA's AI Cyber Challenge~\cite{aixcc}, a competition for developing autonomous cyber reasoning systems (CRSs) that detect and repair vulnerabilities in software.
    Its challenges are carefully constructed by the organizers from mature, widely-used open-source projects: each project is seeded with a set of known vulnerabilities.
    From these, we use all full-scan challenges (whole-repo scan tasks that match our setting) and retain the vulnerabilities with a reliably reproducing PoV, yielding 39 vulnerabilities across 9 projects spanning C, C++, and Java.
    Except for \texttt{nginx}, which served as a demonstration challenge, all challenges were publicly released in May~2026.

    \item \textbf{The OSV benchmark}~(Table~\ref{tab:osv}) is a living, rotating set of recently disclosed vulnerabilities in widely-used open-source projects that we assemble from the OSV database~\cite{osv}.
    We use OSV rather than the more widely used CVE database because each OSV record ties a vulnerability to its introducing and fixing commits and to a public PoV with the disclosed crash information---the per-bug provenance a reproducible benchmark needs but CVE rarely provides.
    To mitigate data-leakage risk, we restrict the snapshot to the most recently disclosed records, covering February--April~2026.
    Of the $45$ vulnerabilities disclosed in this window, only $29$ provided a public PoV at the time of collection, of which $24$ reproduce reliably---yielding $24$ confirmed bugs across $9$ projects spanning C, C++, Java, and Rust.
\end{enumerate}

\textbf{Comparison Tools.}
We compare \tool against agent baselines that span the design space of our approach:

\begin{enumerate}[leftmargin=*]
    \item \textbf{\claude}~\cite{claudecode}, Anthropic's frontier general-purpose coding agent, represents the minimally-structured harness in the design space.
    It audits a codebase by reasoning over it directly.
    It can also invoke tools, e.g., code execution, to ground its analysis, but does so in a completely \emph{ad-hoc} manner. 
    Run under the same model as \tool, \claude demonstrates the utility of \tool's more structured, specification-first approach over ad-hoc agentic reasoning alone.
    \item \textbf{\atlantis}~\cite{atlantiscrs, kim2025atlantis, atlantispaper}, the cyber reasoning system from Team Atlanta that won the AIxCC final, combines the same two ingredients~(an LLM agent and a fuzzer) as \tool.
    However, in contrast to \tool, \atlantis represents a fuzzing-centric approach: an ensemble of fuzzers and concolic executors forms its base infrastructure, while LLM agents support the fuzzing campaign.
    These LLM-based components serve as a semantic front-end that feeds and steers fuzzing, including generating format-aware seeds, selecting fuzzing targets, and synthesizing PoVs for sinks the fuzzer reaches but cannot trigger.
    Comparing against \atlantis thus isolates the contribution of our infer-falsify-refine loop over supplementing a fuzzer with agentic capabilities.
\end{enumerate}

\begin{table*}[h]
\centering
\caption{Bug-finding Results on the AIxCC Benchmark for different tools with different backing models.}
\resizebox{\textwidth}{!}{
\begin{tabular}{ll|OOBB|OOBB|OOBB}
 \toprule
  \textbf{Project}       & \textbf{Challenges}               & \multicolumn{4}{c|}{\textbf{\tool}}              & \multicolumn{4}{c|}{\textbf{\atlantis}} & \multicolumn{4}{c}{\textbf{\claude}} \\
  \cmidrule(lr){3-6}\cmidrule(lr){7-10}\cmidrule(lr){11-14}
  & & \multicolumn{2}{c}{\claudelogo\,Claude} & \multicolumn{2}{c|}{\deepseeklogo\,DeepSeek} & \multicolumn{2}{c}{\claudelogo\,Claude} & \multicolumn{2}{c|}{\deepseeklogo\,DeepSeek} & \multicolumn{2}{c}{\claudelogo\,Claude} & \multicolumn{2}{c}{\deepseeklogo\,DeepSeek} \\
  \rowcolor{white} & & Existing & New & Existing & New & Existing & New & Existing & New & Existing & New & Existing & New  \\
  \hline
  \hline
\texttt{nginx} & \texttt{challenge-04\_1}          & 8/11 & 4     & 3/11   & 1  & 6/11 & 0 & 4/11 & 1 & 6/11 & 1 & 4/11  & 2 \\
\texttt{nginx} & \texttt{challenge-04\_2}          & 1/2  & 0     & 2/2   & 1  & 2/2  & 0 & 2/2 & 0 & 2/2  & 0 & 2/2  & 0 \\
\texttt{nginx} & \texttt{challenge-04\_3}          & 1/1  & 0     & 1/1   & 2  & 1/1  & 0 & 1/1 & 0 & 1/1  & 0 & 1/1  & 0 \\
\texttt{dav1d} & \texttt{dav1d-001}                & 0/1  & 0     & 1/1  & 0   & 1/1 & 0 & 1/1 & 0 & 1/1 & 0 & 1/1 & 1  \\
\texttt{little-cms} & \texttt{lcms-001}            & 1/1  & 0     & 1/1  & 0   & 0/1 & 0 & 0/1 & 0 & 1/1 & 0 & 0/1 & 0  \\
\texttt{little-cms} & \texttt{lcms-002}            & 0/1  & 0     & 0/1  & 0   & 0/1 & 0 & 0/1 & 0 & 0/1 & 0 & 0/1 & 0  \\
\texttt{mongoose} & \texttt{mongoose\_0}           & 1/1  & 1     & 1/1  & 0   & 1/1 & 0 & 1/1 & 0 & 1/1 & 0 & 1/1 & 1  \\
\texttt{apache-poi} & \texttt{vuln\_0,vuln\_1}     & 2/2  & 1     & 2/2  & 1   & 2/2 & 0 & 2/2 & 1 & 2/2 & 6 & 2/2 & 1  \\
\texttt{apache-poi} & \texttt{vuln\_2}             & 1/1  & 8     & 1/1  & 5   & 1/1 & 9 & 0/1 & 10 & 1/1 & 1 & 0/1 & 0  \\
\texttt{apache-poi} & \texttt{vuln\_3}             & 1/1  & 0     & 1/1  & 1   & 0/1 & 0 & 0/1 & 0 & 1/1 & 1 & 1/1 & 0  \\
\texttt{apache-poi} & \texttt{vuln\_4}             & 1/1  & 1     & 0/1  & 0   & 0/1 & 1 & 0/1 & 0 & 1/1 & 0 & 1/1 & 0  \\
\texttt{shadowsocks} & \texttt{libev\_0--4}        & 5/5  & 1     & 5/5  & 2   & 1/5 & 0 & 1/5 & 0 & 5/5 & 0 & 5/5 & 0  \\
\texttt{systemd} & \texttt{systemd-001}            & 1/1  & 0     & 1/1  & 1   & 1/1 & 0 & 1/1 & 0 & 1/1 & 0 & 1/1 & 0  \\
\texttt{systemd} & \texttt{systemd-003}            & 1/1  & 0     & 1/1  & 1   & 1/1 & 0 & 1/1 & 0 & 1/1 & 0 & 1/1 & 0  \\
\texttt{systemd} & \texttt{systemd-004}            & 1/1  & 0     & 1/1  & 0   & 1/1 & 0 & 1/1 & 0 & 1/1 & 0 & 1/1 & 0  \\
\texttt{systemd} & \texttt{systemd-005}            & 1/1  & 2     & 1/1  & 2   & 0/1 & 0 & 0/1 & 0 & 0/1 & 0 & 1/1 & 0  \\
\texttt{wireshark} & \texttt{vuln\_001}            & 1/1  & 0     & 1/1  & 0   & 1/1 & 0 & 1/1 & 0 & 1/1 & 0 & 1/1 & 0  \\
\texttt{wireshark} & \texttt{vuln\_002}            & 1/1  & 5     & 1/1  & 2   & 1/1 & 1 & 0/1 & 2 & 1/1 & 1 & 1/1 & 1  \\
\texttt{wireshark} & \texttt{vuln\_005}            & 1/1  & 1     & 1/1  & 1   & 1/1 & 0 & 1/1 & 0 & 1/1 & 2 & 1/1 & 0  \\
\texttt{wireshark} & \texttt{vuln\_010}            & 1/1  & 0     & 1/1  & 0   & 1/1 & 0 & 1/1 & 0 & 1/1 & 0 & 1/1 & 0  \\
\texttt{wireshark} & \texttt{vuln\_011}            & 1/1  & 0     & 1/1  & 0   & 1/1 & 0 & 1/1 & 0 & 1/1 & 0 & 1/1 & 0  \\
\texttt{wireshark} & \texttt{vuln\_012}            & 1/1  & 1     & 1/1  & 1   & 1/1 & 0 & 1/1 & 0 & 1/1 & 0 & 1/1 & 1  \\
\texttt{xz} & \texttt{xz-001}                     & 1/1  & 1     & 1/1  & 1   & 1/1 & 0 & 1/1 & 0 & 1/1 & 0 & 1/1 & 0  \\
\hline
\textbf{Subtotal}  & & 33/39 & 26 & 29/39 & 22 & 25/39 & 11 & 21/39 & 14 & 32/39 & 12 & 29/39 & 7 \\
\hline \hline
\multicolumn{2}{l|}{\textbf{Total (Existing+New)}} & \multicolumn{2}{>{\columncolor{claudebg}}c}{\textbf{59}} & \multicolumn{2}{>{\columncolor{deepseekbg}}c|}{\textbf{51}} & \multicolumn{2}{>{\columncolor{claudebg}}c}{\textbf{36\impr{63}}} & \multicolumn{2}{>{\columncolor{deepseekbg}}c|}{\textbf{35\impr{45}}} & \multicolumn{2}{>{\columncolor{claudebg}}c}{\textbf{44\impr{34}}} & \multicolumn{2}{>{\columncolor{deepseekbg}}c}{\textbf{36\impr{41}}} \\
\bottomrule
 \end{tabular}
    }
\resultsnote
\label{tab:aixcc}
\end{table*}

\textbf{Configurations.}
We run every system through OSS-CRS~\cite{chin2026oss}, the open-source orchestration framework for evaluating Cyber Reasoning Systems~(CRSs) maintained under the OpenSSF. A CRS performs both bug detection and repair. We evaluate only the detection task that \tool targets. OSS-CRS runs any tool against an OSS-Fuzz-format target via a designated test harness, as in the AIxCC competition~\cite{aixcc}, which omits the false positive issues since all submitted PoVs are through a trusted entry point.%

\emph{Resources and Models}.
Following the AIxCC semifinal budget~\cite{aixccrules}, each system runs for \SI{4}{\hour} per challenge with up to \$100 in LLM API credits, in a Docker container capped at $12$ CPU cores and $32$\,GB of RAM. All experiments run on an Intel~Xeon~Platinum~8468V server (96 cores, 512\,GB of RAM). To separate the effect of the underlying LLM from that of the tool built around it, we run each agentic system under two frontier LLMs: the \emph{proprietary} Claude~Sonnet~4.6~\cite{claude-sonnet-4-6} and the \emph{open-weight} DeepSeek~V4~Pro~\cite{deepseek-v4-pro} (we use ``Claude'' and ``DeepSeek'' for short hereafter).

\emph{Prompt for Baselines}.
We run both baselines exactly as packaged in OSS-CRS, rather than tuning their prompts ourselves, so that neither is advantaged or disadvantaged by our setup.
For \claude, we use the framework's bug-finding Claude Code agent~\cite{claudecodecrs}, prompted to audit the target and produce proof-of-vulnerability inputs for the given test harness. For \atlantis, we reuse the competition configuration its authors contributed to the framework.

\subsection{RQ1: Reproducing Known Vulnerabilities}
\label{sec:eval:rq1}

We compare \tool with the two baselines on both benchmarks.
For each tool, we measure the two quantities reported in Tables~\ref{tab:aixcc} and~\ref{tab:osv}: (1) ``\emph{Existing}'', the number of known, ground-truth vulnerabilities it reproduces, i.e., its recall against a reproducible PoV; and (2) ``\emph{New}'', the number of unique additional bugs found on those subjects at their benchmarked versions, de-duplicated by top-3 stack trace locations.
We report the results of each agentic system under both Claude and DeepSeek.
Because DeepSeek~V4~Pro is substantially cheaper to run, we repeat each of its runs four times and report the average.
All tools see the same projects, versions, entry points, and per-subject budget (\secref{sec:eval_setup}), and every finding is confirmed by the same reproduction-and-deduplication procedure. 

    \begin{table*}[h]
    \centering
    \caption{Bug-finding Results on the OSV Benchmark for different tools with different backing models 
    (``--'' indicates the tool failed to run on the given project)
    }
    \resizebox{\textwidth}{!}{
    \begin{tabular}{ll|OOBB|OOBB|OOBB}
      \toprule
       \textbf{Project}       & \textbf{Commit} & \multicolumn{4}{c|}{\textbf{\tool}}              & \multicolumn{4}{c|}{\textbf{\atlantis}} & \multicolumn{4}{c}{\textbf{\claude}} \\
       \cmidrule(lr){3-6}\cmidrule(lr){7-10}\cmidrule(lr){11-14}
       & & \multicolumn{2}{c}{\claudelogo\,Claude} & \multicolumn{2}{c|}{\deepseeklogo\,DeepSeek} & \multicolumn{2}{c}{\claudelogo\,Claude} & \multicolumn{2}{c|}{\deepseeklogo\,DeepSeek} & \multicolumn{2}{c}{\claudelogo\,Claude} & \multicolumn{2}{c}{\deepseeklogo\,DeepSeek} \\
       \rowcolor{white} & & Existing & New & Existing & New & Existing & New & Existing & New & Existing & New & Existing & New  \\
       \hline \hline
    
    \texttt{apache-poi}    & \texttt{eafd6c0}& 0/2 & 1 & 0/2 & 0 & 0/2 & 0 & 0/2 & 0 & 0/2 & 0 & 0/2 & 0 \\
    \texttt{assimp}        & \texttt{21607df}& 0/1 & 4 & 0/1 & 11 & 1/1 & 19 & 0/1 & 2 & 0/1 & 24 & 0/1 & 7 \\
    \texttt{gpsd}          & \texttt{4f56109}& 1/1 & 3 & 1/1 & 7 & 1/1 & 0 & 1/1 & 0 & 1/1 & 0 & 1/1 & 0 \\
    \texttt{gpsd}          & \texttt{d700650}& 0/3 & 11 & 1/3 & 7 & 0/3 & 0 & 0/3 & 0 & 0/3 & 0 & 0/3 & 0 \\
    \texttt{gpsd}          & \texttt{50153e3}& 0/4 & 11 & 0/4 & 3 & 0/4 & 0 & 0/4 & 0 & 0/4 & 0 & 0/4 & 1 \\
    \texttt{gpsd}          & \texttt{1c9dd87}& 1/3 & 12 & 1/3 & 5 & 1/3 & 0 & 1/3 & 0 & 2/3 & 0 & 1/3 & 0 \\
    \texttt{pjsip}         & \texttt{2b7c8b5}& 1/1 & 0 & 1/1 & 1 & 0/1 & 0 & 0/1 & 0 & 1/1 & 0 & 1/1 & 0 \\
    \texttt{libhevc}       & \texttt{6763519}& 0/1 & 0 & 0/1 & 1 & 1/1 & 0 & 1/1 & 0 & 0/1 & 0 & 0/1 & 0 \\
    \texttt{libical}       & \texttt{d6f6b1c}& 2/3 & 1 & 2/3 & 1 & 3/3 & 0 & 3/3 & 0 & 3/3 & 1 & 1/3 & 1 \\
    \texttt{libical}       & \texttt{fe0de01}& 1/2 & 2 & 1/2 & 2 & 1/2 & 1 & 1/2 & 0 & 2/2 & 0 & 2/2 & 2 \\
    \texttt{tinyobjloader} & \texttt{7dbc543}& 1/1 & 2 & 1/1 & 1 & 1/1 & 0 & 1/1 & 0 & 1/1 & 0 & 1/1 & 0 \\
    \texttt{wasmtime}      & \texttt{d54924a}& 0/1 & 0 & 0/1 & 0 & -- & -- & -- & -- & 0/1 & 0 & 1/1 & 0 \\
    \texttt{mongoose}      & \texttt{ba76869}& 1/1 & 3 & 1/1 & 1 & -- & -- & -- & -- & 0/1 & 1 & 0/1 & 0 \\
    \hline
    \textbf{Subtotal}  & & 8/24 & 50 & 9/24 & 40 & 9/22 & 20 & 8/22 & 2 & 10/24 & 26 & 8/24 & 11 \\
    \hline \hline
    \multicolumn{2}{l|}{\textbf{Total (Existing+New)}} & \multicolumn{2}{>{\columncolor{claudebg}}c}{\textbf{58}} & \multicolumn{2}{>{\columncolor{deepseekbg}}c|}{\textbf{49}} & \multicolumn{2}{>{\columncolor{claudebg}}c}{\textbf{29\impr{86}}} & \multicolumn{2}{>{\columncolor{deepseekbg}}c|}{\textbf{10\impr{370}}} & \multicolumn{2}{>{\columncolor{claudebg}}c}{\textbf{36\impr{61}}} & \multicolumn{2}{>{\columncolor{deepseekbg}}c}{\textbf{19\impr{157}}} \\
    \bottomrule
      \end{tabular}
    }
    \resultsnote
    \label{tab:osv}
    \end{table*}

\textbf{AIxCC Benchmark Results.}
Table~\ref{tab:aixcc} presents the results on the AIxCC benchmark.
On average, \tool finds 34--63\% more bugs than either \atlantis or \claude, and this advantage holds under both Claude and DeepSeek.
On the existing bugs already known to the competition, \tool is comparable with \claude and better than \atlantis.
But for new bugs, \tool finds more than \emph{twice} as many vulnerabilities as either baseline.
Moreover, \tool finds these new bugs across more subjects than the baselines: combining both models, \tool uncovers new bugs on 16 of the 23 challenge groups, against 9 for \claude and 5 for \atlantis.
Its advantage therefore reflects broad effectiveness rather than a few favorable targets.

\textbf{OSV Benchmark Results.}
Table~\ref{tab:osv} reports the results on the OSV benchmark.
\atlantis fails to run on \texttt{mongoose} and \texttt{wasmtime} due to configuration issues on that version and lack of Rust support, respectively.
Overall, \tool finds 61--370\% more bugs than the baselines.
As on the AIxCC benchmark, the vast majority of these additional bugs were \emph{not previously known} in the benchmark, and \tool again finds them more broadly than the baselines, uncovering new bugs in 12 of the 13 commits, against 5 for \claude and 3 for \atlantis.
Notably, all of these programs are kept under \emph{continuous fuzzing} via OSS-Fuzz~\cite{ossfuzz}, so the bugs are \emph{non-trivial}, having evaded detection at the time of benchmark construction.

\textbf{Cross-Model / Agent Comparison.}
Looking at both Table~\ref{tab:aixcc} and Table~\ref{tab:osv}, we can see that in 10/12 subtotal configurations, agents backed by Claude outperform or are at parity with the same agent backed by DeepSeek, with the lone exception being that \atlantis backed by DeepSeek finds 3 more \emph{new} bugs on the AIxCC benchmark.
Importantly, \tool using DeepSeek finds more total vulnerabilities on average than either \atlantis or \claude using Claude across both benchmarks.
In other words, \tool's specification-first approach can leverage an \emph{open-weight model} more effectively than other agents can leverage a \emph{frontier} model.
We believe this result is of increased importance with recent public attention on expensive or limited availability specialized frontier models, such as Claude Mythos~\cite{mythos}.

\subsection{RQ2: Component Contribution}
\label{sec:eval:rq2}

To better understand \tool{}'s effectiveness, we measure each component's contribution by attributing every confirmed vulnerability \tool finds on both benchmarks (\secref{sec:eval:rq1}) to the component that \emph{directly} surfaced it. Each finding falls into one of three discovery lanes: (1)~\emph{invariant falsification}, where the fuzzer breaks an agent-committed invariant, guiding the agent to analyze the resulting violation and trace it to a bug; 
(2)~\emph{fuzzing}, where the fuzzer triggers an ordinary sanitizer crash without violating an invariant; 
and 
(3)~\emph{code review}, where the agent flags a suspect while reviewing the code and reports it once validated, the mechanism that conventional agentic bug scanners rely on.

\begin{table}[t]
\centering
\caption{\tool's exposed vulnerabilities, broken down by discovery lane for each benchmark and backing LLM. The three lane counts in each row sum to \emph{Total}.}
\resizebox{\linewidth}{!}{
\begin{tabular}{ll|r|rrr}
\toprule
\textbf{Benchmark} & \textbf{LLM} & \textbf{Total} & \textbf{Invariant} & \textbf{Fuzzing} & \textbf{Code Review} \\
\midrule
\multirow{2}{*}{AIxCC} & \claudelogo\,Claude & 59   & 21         & 9            & 29            \\
                       & \deepseeklogo\,DeepSeek & 51   & 17           & 10            & 24            \\
\midrule
\multirow{2}{*}{OSV}   & \claudelogo\,Claude & $58$ & $24$ & $9$ & $25$ \\
                       & \deepseeklogo\,DeepSeek & $49$    & $13$            & $13$            & $23$      \\

\bottomrule
\end{tabular}%
}
\label{tab:discovery_lanes}
\end{table}

\textbf{Results.}
Table~\ref{tab:discovery_lanes} shows that all three discovery lanes contributed to \tool's success on both benchmarks.
Fuzzing alone was able to surface a small number of relatively shallow bugs on both benchmarks.
However, across backing LLMs, a \emph{large} number of vulnerabilities can be detected using only \emph{static reasoning} via code review.
This finding supports recent anecdotal evidence that scanning code with AI agents can already identify many new vulnerabilities~\cite{grinstead_holler_braun_2026}.

Despite the effectiveness of code review, we also see that inferred invariants directly lead to a large number of vulnerability discoveries.
Indeed, nearly half (36-41\%) of all vulnerabilities discovered by \tool with Claude were found via invariant trips.
However, with DeepSeek, this number decreases substantially.
Indeed, \emph{nearly all} of the drop in effectiveness between \tool using Claude and DeepSeek as backing models on the OSV benchmark can be attributed to the fewer bugs found via invariant trips.
On manual inspection, we found that DeepSeek struggles to maintain focus on the longer horizon task of setting and refining invariants.
A possible future line of research could be into more sophisticated context management to alleviate this issue for open-weight models.

We also examine the auditing cost. Averaged over both benchmarks, an audit costs {(\$45.41, \$2.19)} for \tool, compared to {(\$73.90, \$3.20)} for \atlantis and {(\$16.04, \$0.30)} for \claude, under the form of ``(Claude~Sonnet~4.6~\cite{claude-sonnet-4-6}, DeepSeek~V4~Pro~\cite{deepseek-v4-pro})''.
Unlike \tool and \atlantis, the default \claude harness does not record exact pricing, so we compute its cost from token usage in the session transcripts, de-duplicated by message ID and priced at published provider rates.
We see that \tool is similar in cost to alternatives \emph{and} that it gives comparable results on the cheaper DeepSeek to those on the more expensive Claude across our two benchmarks.
Moreover, \tool with DeepSeek already exceeds the capabilities of baselines backed by Claude at a fraction of the cost!
Moreover, as the inferred invariants are committed into the source as durable instrumentation, much of this cost is a one-time investment that can be amortized over later audits.

\begin{table*}[t]
\centering
\caption{Summary of previously unknown vulnerabilities found by \tool}
\label{tab:newbugs}
\resizebox{\linewidth}{!}{
\begin{tabular}{clll}
\toprule
\textbf{\#} & \textbf{Project} & \textbf{Bug Description} & \textbf{Status} \\
\midrule
1  & \texttt{chisel}   & ACL bypass in User.HasAccess() when regexp.MatchString() evaluates unanchored ACL patterns & Confirmed \\
2  & \texttt{chisel}   & Access-control bypass in \texttt{tunnel\_out\_ssh.go} due to the unchecked host:port in a post-handshake channel's \texttt{ExtraData()} & Fixed \\
3  & \texttt{Ghost}    & \texttt{LocalStorageBase.save()} ignores type parameter, serving uploaded \texttt{.html} as \texttt{text/html} & Reported \\
4  & \texttt{Ghost}    & SSRF in \texttt{Webhook.trigger()} and Slack via \texttt{@tryghost/request} skipping \texttt{request-external.js} IP checks & Reported \\
5  & \texttt{gpsd}     & RTK baseline data in \texttt{processPSTI030()}/\texttt{processPSTI032()} silently discarded when PSTI opens a new cycle & Fixed \\
6  & \texttt{gpsd}     & Heap buffer overflow in \texttt{ais\_binary\_decode()} when a Type 21 message appends a 16-char name extension past \texttt{name[35]} & Fixed \\
7 & \texttt{gpsd}     & Out-of-bounds read in \texttt{decode\_itk\_pseudo()} on a malformed packet whose length check only warns instead of returning & Fixed \\
8 & \texttt{gpsd}     & One-byte buffer overflow in \texttt{ais\_binary\_decode()} when an un-padded 1008-bit Type 14 text writes past \texttt{text[161]} & Fixed \\
9 & \texttt{gpsd}     & Pointer write past \texttt{field[]} in NMEA field splitting when a sentence has more comma-separated fields than \texttt{NMEA\_MAX\_FLD} & Fixed \\
10 & \texttt{gpsd}     & Signed left-shift undefined behavior in the driver-identification bitmask (\texttt{1 << driver\_index}) in \texttt{libgpsd\_core.c} & Fixed \\
11 & \texttt{gpsd}     & Out-of-bounds write past \texttt{sats\_used[]} in TSIP \texttt{decode\_x6c()} when the 8-bit \texttt{count} field exceeds \texttt{MAXCHANNELS} & Fixed \\
12 & \texttt{gpsd}     & Out-of-bounds read in the RTCM3 CRC-failure path when logging indexes \texttt{inbufptr} instead of \texttt{inbuffer} in \texttt{packet.c} & Reported \\
13 & \texttt{gpsd}     & Out-of-bounds write past \texttt{skyview[]} in the Skytraq \texttt{0xDE} decoder when records exceed \texttt{MAXCHANNELS} before the post-loop clamp & Fixed \\
14 & \texttt{gpsd}     & Source-side out-of-bounds read in the RTCM3 1008/1033 string decoders when later length fields skip cumulative bounds checks & Fixed \\
15 & \texttt{gpsd}     & Out-of-bounds read past \texttt{skyview[]} in the Skytraq \texttt{0xDE} decoder when \texttt{skyview[st].used} is read after \texttt{st} is incremented & Fixed \\
16 & \texttt{lightway} & \texttt{try\_from\_wire()} advances the anti-replay window by plaintext \texttt{wire\_counter} before AES-GCM auth, dropping valid packets & Fixed \\
17 & \texttt{lightway} & \texttt{flags} field excluded from the AES-GCM AAD, so flipping its \texttt{encoded} bit reroutes the packet into a fatal decoder error & Fixed \\
18 & \texttt{lightway} & Plaintext \texttt{version} mismatch in \texttt{outside\_data\_received()} is treated as fatal, so one packet disconnects the client & Fixed \\
19 & \texttt{ntpd-rs}  & IPv4-mapped IPv6 filter bypass in \texttt{IpFilter::is\_in()} on dual-stack sockets, skipping IPv4 allowlists rules & Reported \\
20 & \texttt{rack}     & Regex injection in \texttt{Rack::Sendfile} caused by interpolating the unescaped \texttt{X-Accel-Mapping} header into a path-rewrite regex & Fixed \\
21 & \texttt{zlib}     & Integer overflow in \texttt{zipOpenNewFileInZip4\_64()} when the 16-bit extra-field guard omits the appended 20-byte ZIP64 block & Reported \\
22 & \texttt{zlib}     & NULL dereference in \texttt{zipOpenNewFileInZip4\_64()} as ALLOC'd \texttt{central\_header} is written for 40+ lines before NULL check & Reported \\
\bottomrule
\end{tabular}%
}
\end{table*}

\subsection{RQ3: Detecting Unknown Vulnerabilities}
\label{sec:eval:rq3}

We next assess whether \tool finds previously unknown vulnerabilities in the wild, beyond the two benchmarks above. 
We ran \tool on a set of widely-used and well-tested open-source projects that together span multiple programming languages and application domains, from network protocols and daemons to web applications and libraries.
Here we follow the same setup as our benchmark experiments~(\secref{sec:eval_setup}), with two relaxations suited to a real-world audit rather than a controlled benchmark comparison: \tool is given no designated test harness or entry point, and a finding need not trigger a crash but may instead be a logic flaw. 
\tool itself attempts to synthesize a working PoV for each candidate. 
For memory-safety bugs this PoV is a crashing input, whereas for logic and access-control flaws it is an input that violates a high-level security property outlined in the threat model, but may not necessarily crash the program.
Because such findings may include logic flaws that no sanitizer can flag, we validated each candidate manually, reading \tool's report and any PoV it produced, tracing the issue through the source code, and judging whether the behavior genuinely violates the developers' intent. 
We count a finding only if it is previously unknown, has no public report at audit time, and passes this manual validation.
We disclosed every confirmed finding to its maintainers under the responsible disclosure policy. 

\textbf{Results.}
Table~\ref{tab:newbugs} lists the \bugnum previously unknown vulnerabilities \tool discovered across seven widely-used open-source projects. Their maintainers have already fixed or confirmed \fixedbugnum of them, and two have been assigned CVEs~(CVE-2026-48113 for Bug~\#2 in \texttt{chisel} and CVE-2026-34830 for Bug~\#20 in \texttt{rack}).
Remaining reports are undergoing CVE assignment.
Due to their severity, the findings on \texttt{lightway} have also earned bug bounty rewards totaling \$1{,}400.
These are all mature, widely-used projects. Some of them, such as \texttt{zlib} and \texttt{gpsd}, are even kept under continuous fuzzing in OSS-Fuzz~\cite{ossfuzz}, so surfacing new bugs in them is non-trivial.
Meanwhile, the findings span diverse vulnerability classes, from classic memory-safety bugs (e.g., the heap-buffer-overflow in \texttt{gpsd}'s \texttt{ais\_binary\_decode()}) to logic and access-control flaws such as \texttt{chisel}'s post-handshake ACL bypass and \texttt{ntpd-rs}'s IPv4-mapped filter bypass. 
Through manual inspection, we further find that many of these findings hinge on counter-intuitive inputs that an agent reasoning over the code alone is unlikely to anticipate. For example, Bug~\#7 requires a malformed packet that slips past a length check which only warns instead of returning, while Bug~\#21 requires a ZIP64 extra field whose 20-byte block a 16-bit size guard silently omits.
Such inputs are hard for either paradigm alone: a purely analytical agent is likely to overlook such counter-intuitive inputs, while plain fuzzing struggles to reach the deep states behind them. \tool makes them far more reachable by turning the agent's invariant into a \emph{shallow, semantic} target the fuzzer can reach and falsify, rather than a deep crash it must stumble upon.

\subsection{RQ4: Usefulness of Specifications}
\label{sec:eval:rq4}

Invariants inferred by \tool are not only a one-time effort but durable artifacts that remain useful long after the audit that produced them. 

We illustrate this with a case study (Listing~\ref{lst:gpsd} and \ref{lst:gpsd-fix}) on \texttt{gpsd}, a deployed GPS daemon and a subject from the OSV benchmark.
In Listing~\ref{lst:gpsd}, \texttt{satellites\_visible} is the number of satellites currently in view, and \texttt{MAXCHANNELS} is the receiver's channel capacity~(lines~\ref{ln:max}--\ref{ln:cnt}). A receiver cannot report more satellites than it has channels, so during its audit \tool infers and records this semantic bound as the in-source invariant $\phi$: \texttt{satellites\_visible} $\le$ \texttt{MAXCHANNELS}~(line~\ref{ln:inv}). On that basis it deems the downstream operations secure, including the 10+ \emph{consumer} functions that walk \texttt{skyview} indexed by \texttt{satellites\_visible}~(lines~\ref{ln:cons}--\ref{ln:consend}).
However, this verdict is fragile: \texttt{satellites\_visible} also has 10+ \emph{producer} drivers that assign it by directly copying raw packet fields~(lines~\ref{ln:prod}--\ref{ln:prodend}), so a crafted packet can drive it to a value such as $256$ and falsify $\phi$, which is how \tool surfaces the bug based on fuzzing's falsification.

\begin{lstlisting}[float=tb, style=cstyle, frame=single, numbers=left, numberstyle=\tiny\color{gray}, numbersep=5pt, xleftmargin=1.2em, linebackgroundcolor={\ifnum\value{lstnumber}<4 \color{workflowblue}\else\ifnum\value{lstnumber}<7 \color{workfloworange}\else\color{workflowgray}\fi\fi}, caption={The \texttt{satellites\_visible} bug family in \texttt{gpsd}, shaded by role: \colorbox{workflowblue}{the invariant $\phi$ over variables}, the \colorbox{workfloworange}{producers}, and the \colorbox{workflowgray}{consumers}.}, label={lst:gpsd}]
#define MAXCHANNELS 184      // a receiver's channel capacity (*@\label{ln:max}@*)
int satellites_visible;      // number of satellites currently in view (*@\label{ln:cnt}@*)
// (*@\underline{\tool{} inferred invariant $\phi$:}\textbf{ satellites\_visible <= MAXCHANNELS}@*)  (*@\label{ln:inv}@*)
// Producers: 10+ drivers set the satellites_visible from raw packet:
geostar(): satellites_visible = (int)getleu32(buf, ...); (*@\label{ln:prod}@*)
sirf() (*@\ @*)    : satellites_visible = num_of_sats; (*@\label{ln:prodend}@*)
// Consumers: 10+ readers walk skyview using satellites_visible:
struct satellite_t skyview[MAXCHANNELS]; (*@\label{ln:cons}@*)
fill_dop()      (*@\quad\quad@*) : for (k=0;k<satellites_visible;k++) ... skyview[k] (*@\label{ln:filldop}@*)
json_sky_dump() : for (i=0;i<satellites_visible;i++) ... skyview[i] (*@\label{ln:consend}@*)
\end{lstlisting}

\begin{lstlisting}[float=tb, style=diffstyle, caption={The four-month remediation history of a bug family in \texttt{gpsd}: (1)~clamping one crash site, an \emph{incomplete fix}; (2)~bounding the value at its multiple sources, fixing the whole \emph{bug family} by enforcing $\phi$; (3)~enlarging the capacity, repairing the {undersized bound}. Across 4 months and 3 batches of fixes, each restores the one invariant $\phi$ \tool inferred once.}, label={lst:gpsd-fix}]
// (1). (*@\underline{Patch (+1day)}@*): clamp the loop at ONE consumer (incomplete fix) (*@\label{ln:fix-1}@*)
- fill_dop(): for (k=0; k<satellites_visible; k++) (*@\label{ln:fix-loop}@*)
+ fill_dop(): for (k=0; k<min(satellites_visible, MAXCHANNELS); k++) (*@\label{ln:fix-loopend}@*)

// (2). Patches: bound the count at the producer drivers, enforcing (*@$\phi$@*) (*@\label{ln:fix-2}@*)
//    (*@\underline{batch-1 (+5days)}@*): clamp the VALUE at ~10 producer drivers
- driver(): satellites_visible = <count from packet>;
+ driver(): satellites_visible = min(<count from packet>, MAXCHANNELS); (*@\label{ln:fix-2-batch1-end}@*)
//    (*@\underline{batch-2 (+8days..+4months)}@*): the sweep missed TSIP's own skyview[] fills (*@\label{ln:fix-2-batch2-start}@*)
- tsip():   ... skyview[i] = ...             // i ran past MAXCHANNELS
+ tsip():   if (count < TSIP_CHANNELS) { ... skyview[i] = ... } (*@\label{ln:fix-2-batch2-end}@*)

// (3). (*@\underline{closure (+4months)}@*): a driver outgrew 184 (violating (*@$\phi$@*)) but it is legit, indicating the array's capacity was undersized (*@\label{ln:fix-3}@*)
- #define MAXCHANNELS 184
+ #define MAXCHANNELS 230
+ #if MAXCHANNELS < SKY_CHANNELS   // SKY_CHANNELS = 230, SkyTraq's max
+   #error  // the array must cover SkyTraq
+ #endif (*@\label{ln:fix-3end}@*)
\end{lstlisting}

Tracing \texttt{gpsd}'s subsequent development history shows why $\phi$ is \emph{durable}: inferred \emph{once}, it keeps flagging the same root cause across a four-month chain of remediations~(Listing~\ref{lst:gpsd-fix}), first exposing an \textbf{incomplete fix}, then tying a whole \textbf{bug family} to that one cause, and finally revealing that even the bound $\phi$ assumes was itself too low.

\emph{(1) Exposing an incomplete fix.} The family first surfaced as the disclosed report OSV-2026-189~\cite{osv2026189}, an out-of-bounds read in the \texttt{fill\_dop} consumer~(line~\ref{ln:filldop}) that violates $\phi$. The developers responded within a day, but their patch clamped only the loop in that single consumer \texttt{fill\_dop}~(Listing~\ref{lst:gpsd-fix}, lines~\ref{ln:fix-1}--\ref{ln:fix-loopend}). Such a loop-local fix could silence a sanitizer: the lone out-of-bounds access it would catch, in \texttt{fill\_dop}, is now bounded, and the other consumers cannot be reached by this input, so the defect appears resolved. However, re-checking $\phi$ still flags a violation: the producers~(e.g., \texttt{geostar}, line~\ref{ln:prod}) keep storing the raw count, and the consumers the patch missed~(e.g., \texttt{json\_sky\_dump}, line~\ref{ln:consend}) remain unbounded. Because $\phi$ constrains the program state rather than a single access site, it catches this latent corruption even when no out-of-bounds access executes. This is exactly what a sanitizer-guided fix overlooks.

\emph{(2) Tying together a bug family.} A recurrence soon makes the root cause more clear. A second report, a crash reached through the \texttt{tsip} driver, surfaced at a different site. Despite the different symptom, it violates the \emph{same} invariant $\phi$, so the two are not isolated bugs but one root cause shared across a family of producers and consumers. Recognizing this, the developers moved the fix to the \emph{producers}: as shown in Listing~\ref{lst:gpsd-fix}, they bounded the count at its source across roughly ten drivers~(lines \ref{ln:fix-2}--\ref{ln:fix-2-batch1-end}) and extended the sweep to the cases the first pass missed, such as \texttt{tsip}'s own \texttt{skyview} fills~(lines~\ref{ln:fix-2-batch2-start}--\ref{ln:fix-2-batch2-end}). This remediation for such a bug family runs from five days to four months after the first report, and each of them enforces exactly what $\phi$ asserts.

\emph{(3) Revealing an undersized bound.} A later input pushed \texttt{satellites\_visible} past the upper bound \texttt{MAXCHANNELS}, yet triage showed the value to be \emph{legitimate}: a \texttt{SkyTraq} receiver genuinely reports up to 230 satellites, beyond the maximum capacity \texttt{MAXCHANNELS} assumes (i.e., 184). The violation thus pointed not to a producer but to the bound itself.
Upstream consequently enlarged the array to 230, under a compile-time guard ensuring it never falls below \texttt{SkyTraq}'s channel count~(Listing~\ref{lst:gpsd-fix}, lines~\ref{ln:fix-3}--\ref{ln:fix-3end}).

Across all the three episodes, $\phi$ pins down the single property $\phi$ every complete fix had to restore, whether by bounding the value or enlarging the array's capacity, and outlives both the original report and each point fix.
We are currently engaged in active discussions with developers of \texttt{gpsd} to merge this invariant as a defensive check against future regressions.


\section{Related Work}
\label{sec:related}

\textbf{Agentic and LLM Assisted Fuzzing.}
Many works have proposed using LLMs to assist fuzzing in harness generation~\cite{ossfuzzgen, lyu2024prompt, liu2025promefuzz}, seed creation~\cite{luo2025enhancing}, and taint analysis~\cite{ji2026firmagent}.
This was also the strategy used by many teams in the recent DARPA AIxCC~\cite{ atlantispaper, sheng2025fuzzingbrain, wolff2026large}.
Recent works have even begun to replace fuzzers altogether with LLM or agentic approaches~\cite{luo2026concollmic, xia2024fuzz4all}. 
We have compared \tool with the winning AIxCC system.
Additionally, while these approaches are all fuzzing-centric, we take an inverted approach in this work: fuzzing is a tool used by human security analysts and \tool alike to find edge cases and check assumptions.
Indeed, advances in fuzzing can be leveraged by \tool's fuzzing component to increase its effectiveness in invariant falsification.

\textbf{Agentic Vulnerability Scanning.}
Recently, agentic vulnerability scanning systems, such as Claude Mythos~\cite{mythos} and Google's Big Sleep~\cite{bigsleep} have gained widespread attention.
In addition to custom models~\cite{mythos}, several works have explored approaches such as Retrieval Augmented Generation (RAG)~\cite{du2024vulrag}, deep-agents~\cite{park2026agentic}, and role-based multi-agent systems~\cite{wang2025vulagent, wei2025smartmultiagent, hu2023largesmart}.
These systems, however, keep their security reasoning \emph{implicit} within the agent, as discussed in \secref{sec:motivation:limitations}. \tool instead externalizes it as explicit, executable security specifications and continuously refines them via dynamic falsification.

\textbf{Specification Inference.} Specification inference~\cite{ernst2007daikon, lemieux2015general} constitutes deduction of the \emph{intended} behavior of code without explicit formalized descriptions of this behavior.
LLMs have demonstrated impressive capacity to infer intended specifications related to \emph{repairing} buggy code~\cite{specrover}, functional requirements~\cite{mu2024clarifygpt}, and formal specifications~\cite{ma2025specgen, endres2024can, lahirie2024evaluating}.
A key contribution of this work is to likewise examine LLMs' capacity for inferring security specifications for vulnerability \emph{discovery}.
In the context of fuzzing, ECG~\cite{zhang2024ecg} and Halo~\cite{huang2024everything} infer \emph{input specifications} to assist in embedded systems fuzzing and to constrain the input space in directed fuzzing, respectively.
Similarly, Locus~\cite{zhu2025locus} infers specifications to \emph{prune} infeasible code paths in directed fuzzing.
These approaches contrast with the \emph{local invariants} and \emph{guidance} leveraged by \tool.
Fioraldi et al.~\cite{fioraldi2021use} also infer invariants to guide grey-box fuzzing, but do so from the behavior of \emph{observed execution traces}, rather than based on the necessary conditions for code to be secure, as in \tool. 
FM-Agent~\cite{ding2026fmagent} infers pre- and post-conditions for functions in large software projects and uses ad-hoc test-case generation to identify bugs.
IRIS~\cite{li2024llm} and LLift~\cite{li2024enhancing} infer some specifications via LLM to assist in static analysis.
However, rather than inferring specifications alone, \tool leverages comprehensive falsification via fuzzing to \emph{validate and refine} its inferred specifications.


\section{Limitations}

\textbf{Reliability of LLMs.}
Agentic vulnerability detection is limited by the inherent unsoundness of LLMs (i.e., the agent may misunderstand the code or even hallucinate).
Our design combats such unsoundness in two ways:
(1)~\emph{reasoning externalization and validation}: we turn the agent's implicit reasoning into explicit, \emph{executable} invariants, an interface that (\romannumeral1)~grounded tools can check, a process less susceptible to omissions or other failures in LLM reasoning itself, and that (\romannumeral2)~developers can also directly read;
(2)~\emph{output validation}: \tool reports a vulnerability only with a generated proof-of-vulnerability input that reproduces it on the original program.
This reason-falsify-refine loop makes \tool more resilient to its own reasoning flaws, helping uncover more bugs as shown in our evaluation.

We did not evaluate \tool in adversarial settings, where an attacker already has sufficient access to modify a project's source code and plant a back-door vulnerability.
This work instead focuses on \emph{inadvertent} vulnerabilities, which are far more common than malicious back-doors.
We view detecting adversarially hidden vulnerabilities as an interesting direction for future research.

\textbf{Data Leakage.}
LLMs are trained on enormous corpora that include code from many open-source projects, so data leakage~\cite{roberts2023cutoff, deng2024benchmark}, where evaluation subjects overlap with the training data, threatens the validity of our results.
We mitigate this risk along several fronts.
First, we created a \emph{new} benchmark, consisting only of bugs reported \emph{after} the training cutoff for Claude Sonnet 4.6.
Unfortunately, for the OSV-benchmark, at the time of writing, no reproducible vulnerabilities had yet been recorded in the OSV database~\cite{osv} fully after the training cutoff for DeepSeek V4.
However, we see a strong correlation between the results obtained by both \tool and other agents with DeepSeek and with Claude (\secref{sec:eval}), and thus we believe that the impact of the possible leakage is not significant.
Second, all AIxCC subjects except \texttt{nginx} were publicly released \emph{after} the training cutoffs of both models, so results on this benchmark are uncontaminated.
Finally, \tool found many previously unknown bugs that we reported to developers, which by definition cannot be attributed to data leakage.

\textbf{Model Context Window.}
Both backed LLMs in our evaluation have a context window limited to 1M tokens, which could be exhausted on very large programs.
We mitigate this by designing \tool as a multi-agent system, where sub-agents are passed only the context necessary for their more granular tasks.
We also set an automatic compaction of context to occur at 800k tokens, but we did not observe \tool reaching this limit in any of our experiments, including on \texttt{wireshark} with 2.1 million source lines of code.

\section{Perspectives}
\label{sec:perspectives}

Software security has always been a fragile balance of power between attackers and defenders.
As new tactics and technologies emerge on the offensive side, they must be mitigated by new defenses.
LLM agents have drastically shifted the balance in this equation; attackers with little to no expertise can point agents at open-source repositories and rapidly discover new vulnerabilities.
We believe that merely replicating this workflow as a defensive measure is drastically insufficient: each vulnerability discovered represents only a single exploitable weakness rather than an underlying security principle.
To reach parity, we formulate an approach in this work that extracts generalized knowledge from each security audit in the form of security specifications.
By making these specifications explicit and checkable, \tool achieves high efficacy in finding vulnerabilities and creates reusable artifacts that help future human and automated analysts reason about the security of their systems.
As an increasing share of code is being written and developed by AI, deepened understanding provided by these specifications will be more valuable than ever.
We believe that our specification-first approach can serve as the basis and inspiration for increasing the use of LLM-based security analysis. 

\subsection{Commentary about Claude Mythos}

In this work, we have presented a rather complex agent harness for detecting security vulnerabilities. Our approach differs significantly, in both methodology and philosophy, from curated Large Language Models like Claude Mythos~\cite{mythos}, which have gained recent attention.
\begin{itemize}
\item First and foremost, models like Claude Mythos currently have restricted access. Fable has been released, but its guardrails prevent it from being used for any security research and development. Thus our agent \tool, built on top of regular models like Claude~Sonnet~4.6 and the open-weight DeepSeek~V4~Pro, provides an alternative when curated models like Claude Mythos are not available. In an experiment on a \texttt{wolfSSL} bug~\cite{mythos-wolfssl} originally reported by Mythos, \tool with the DeepSeek~V4~Pro backend also uncovered it, taking 2.25 hours and \$1.39.

\item Secondly, as mentioned earlier, we believe that instead of finding many bugs using LLMs, it is more important to generalize them in the form of specifications. These specifications, when deposited with a software project, can also prevent vulnerabilities from occurring in the future. Thus, they provide an additional form of regression checking, as we show in the case study in \secref{sec:eval:rq4}, which we cannot achieve using bug-finding models like Mythos alone.

\item Last but not least, we strongly believe that \tool's invariant specifications---once deposited and refined---have widespread usages beyond vulnerability detection! Such usages would not be possible with models alone.
The specifications are an important ingredient in systematizing and documenting the informal reasoning in an AI agent. A core problem in software engineering has always been understanding the developer's intent. Indeed, program analysis techniques have been used in automated program repair to extract a formal description of developer intent---both in the pre-LLM and post-LLM era \cite{semfix,specrover}. In the future, if agents are part of software teams, we will similarly need techniques for understanding agent intent, to effectively manage and maintain a software system. Our work envisions and supports this forward-looking perspective of future software engineering practice.
\end{itemize}

\IEEEpeerreviewmaketitle

\section*{Acknowledgments}
This research is supported by the National Research Foundation, Singapore, under its Artificial Intelligence (AI)-for-Science (AI4S) Challenge Grant (Award No. NRF-AI4SCH-2025-0003), called 
"\href{https://ai4pr.github.io}{AI for Program Reasoning}". Any opinions, findings and conclusions or recommendations expressed in this material are those of the author(s) and do not reflect the views of National Research Foundation.

\bibliographystyle{IEEEtran}
\bibliography{ref.bib}

\begin{thebibliography}{10}
\providecommand{\url}[1]{#1}
\csname url@samestyle\endcsname
\providecommand{\newblock}{\relax}
\providecommand{\bibinfo}[2]{#2}
\providecommand{\BIBentrySTDinterwordspacing}{\spaceskip=0pt\relax}
\providecommand{\BIBentryALTinterwordstretchfactor}{4}
\providecommand{\BIBentryALTinterwordspacing}{\spaceskip=\fontdimen2\font plus
\BIBentryALTinterwordstretchfactor\fontdimen3\font minus
  \fontdimen4\font\relax}
\providecommand{\BIBforeignlanguage}[2]{{%
\expandafter\ifx\csname l@#1\endcsname\relax
\typeout{** WARNING: IEEEtran.bst: No hyphenation pattern has been}%
\typeout{** loaded for the language `#1'. Using the pattern for}%
\typeout{** the default language instead.}%
\else
\language=\csname l@#1\endcsname
\fi
#2}}
\providecommand{\BIBdecl}{\relax}
\BIBdecl

\bibitem{mythos}
\BIBentryALTinterwordspacing
N.~Carlini, N.~Cheng, K.~Lucas, M.~Moore, M.~Nasr, V.~Prabhushankar, W.~Xiao
  Hakeem~Angulu, E.~Ben~Asher, J.~Bow, K.~Bradwell, B.~Buchanan, D.~Forsythe,
  D.~Freeman, A.~Gaynor, X.~Ge, L.~Graham, K.~Guru, H.~Lakhani, M.~McNiece,
  M.~Mehrara, R.~Nichol, A.~Pirzada, S.~Porter, A.~Terzis, and K.~Troy,
  ``{Claude Mythos} preview,'' 2026. [Online]. Available:
  \url{https://red.anthropic.com/2026/mythos-preview/}
\BIBentrySTDinterwordspacing

\bibitem{grinstead_holler_braun_2026}
\BIBentryALTinterwordspacing
B.~Grinstead, C.~Holler, and F.~Braun, ``Behind the scenes hardening firefox
  with claude mythos preview,'' May 2026. [Online]. Available:
  \url{https://hacks.mozilla.org/2026/05/behind-the-scenes-hardening-firefox/}
\BIBentrySTDinterwordspacing

\bibitem{specrover}
H.~Ruan, Y.~Zhang, and A.~Roychoudhury, ``{SpecRover}: Code intent extraction
  via {LLMs},'' in \emph{2025 IEEE/ACM 47th International Conference on
  Software Engineering (ICSE)}, 2025.

\bibitem{ma2025specgen}
L.~Ma, S.~Liu, Y.~Li, X.~Xie, and L.~Bu, ``Specgen: Automated generation of
  formal program specifications via large language models,'' in \emph{2025
  IEEE/ACM 47th International Conference on Software Engineering (ICSE)}.\hskip
  1em plus 0.5em minus 0.4em\relax IEEE, 2025, pp. 16--28.

\bibitem{bigsleep}
\BIBentryALTinterwordspacing
B.~S. Team, ``From {Naptime} to {Big Sleep}: Using large language models to
  catch vulnerabilities in real-world code,'' 2024. [Online]. Available:
  \url{https://projectzero.google/2024/10/from-naptime-to-big-sleep.html}
\BIBentrySTDinterwordspacing

\bibitem{claudecode}
{Anthropic}. (2025) {Claude Code}: An agentic coding tool.
  \url{https://claude.com/product/claude-code}.

\bibitem{aixcc}
{DARPA}. (2025) {AI Cyber Challenge (AIxCC)}.
  \url{https://aicyberchallenge.com/}.

\bibitem{osv}
{Google}. (2021) {OSV}: Open source vulnerabilities database and triage
  service. \url{https://github.com/google/osv.dev}.

\bibitem{atlantiscrs}
{Team Atlanta}. (2025) {Atlantis}: Team atlanta's cyber reasoning system for
  the {DARPA AIxCC} final competition.
  \url{https://github.com/Team-Atlanta/aixcc-afc-atlantis}.

\bibitem{gpsd}
\BIBentryALTinterwordspacing
{The GPSd Project}, ``{GPSd}: Put your {GPS} on the net!'' 2026. [Online].
  Available: \url{https://gpsd.io/}
\BIBentrySTDinterwordspacing

\bibitem{Serebryany2012AddressSanitizerAF}
K.~Serebryany, D.~Bruening, A.~Potapenko, and D.~Vyukov, ``{AddressSanitizer}:
  A fast address sanity checker,'' \emph{2012 USENIX Annual Technical
  Conference}, 2012.

\bibitem{miller1990empirical}
B.~P. Miller, L.~Fredriksen, and B.~So, ``An empirical study of the reliability
  of {UNIX} utilities,'' \emph{Communications of the ACM}, vol.~33, no.~12, pp.
  32--44, 1990.

\bibitem{libfuzzer}
{LLVM Project}, ``{libFuzzer}: A library for coverage-guided fuzz testing,''
  \url{https://llvm.org/docs/LibFuzzer.html}.

\bibitem{jazzer}
{Code Intelligence}, ``{Jazzer}: Coverage-guided, in-process fuzzing for the
  {JVM},'' \url{https://github.com/CodeIntelligenceTesting/jazzer}.

\bibitem{pi}
{earendil-works}. (2026) Pi: An {AI} agent toolkit.
  \url{https://github.com/earendil-works/pi/releases/tag/v0.77.0}.

\bibitem{kim2025atlantis}
T.~Kim, H.~Han, S.~Park, D.~R. Jeong, D.~Kim, D.~Kim, E.~Kim, J.~Kim, J.~Wang,
  K.~Kim \emph{et~al.}, ``{ATLANTIS}: {AI}-driven threat localization,
  analysis, and triage intelligence system,'' \emph{arXiv preprint
  arXiv:2509.14589}, 2025.

\bibitem{atlantispaper}
C.~Zhang, Y.~Park, F.~Fleischer, Y.-F. Fu, J.~Kim, D.~Kim, Y.~Kim, Q.~Xu,
  A.~Chin, Z.~Sheng \emph{et~al.}, ``Sok: Darpa's ai cyber challenge (aixcc):
  Competition design, architectures, and lessons learned,'' \emph{Usenix
  Security}, 2026.

\bibitem{chin2026oss}
A.~Chin, D.~Kim, Y.-F. Fu, F.~Fleischer, Y.~Kim, H.~Han, C.~Zhang, B.~J. Lee,
  H.~Zhao, and T.~Kim, ``{OSS-CRS}: Liberating {AIxCC} cyber reasoning systems
  for real-world open-source security,'' \emph{arXiv preprint
  arXiv:2603.08566}, 2026.

\bibitem{aixccrules}
{DARPA}. (2024) {AIxCC} competition: Procedures and scoring guide.
  \url{https://aicyberchallenge.com/wp-content/uploads/2024/06/ASC-Procedures-and-Scoring-Guide-v4.pdf}.

\bibitem{claude-sonnet-4-6}
Anthropic. (2026) {Claude Sonnet} 4.6.
  \url{https://docs.anthropic.com/en/docs/about-claude/models/overview}.

\bibitem{deepseek-v4-pro}
{DeepSeek-AI}. (2026) {DeepSeek V4 Pro}. \url{https://api-docs.deepseek.com}.

\bibitem{claudecodecrs}
{Team Atlanta}. (2026) {Claude Code} bug-finding agent
  (\texttt{crs-bug-finding-claude-code}).
  \url{https://github.com/Team-Atlanta/crs-bug-finding-claude-code}.

\bibitem{ossfuzz}
K.~Serebryany, ``{OSS-Fuzz}: Google's continuous fuzzing for open-source
  software.''\hskip 1em plus 0.5em minus 0.4em\relax Vancouver, BC: {USENIX}
  Association, Aug 2017.

\bibitem{osv2026189}
{OSV}. (2026) {OSV-2026-189}: Out-of-bounds read in {gpsd}.
  \url{https://osv.dev/vulnerability/OSV-2026-189}.

\bibitem{ossfuzzgen}
\BIBentryALTinterwordspacing
Google, ``{OSS-Fuzz-Gen}: {LLM} powered fuzzing via {OSS-Fuzz},'' 2024.
  [Online]. Available: \url{https://github.com/google/oss-fuzz-gen}
\BIBentrySTDinterwordspacing

\bibitem{lyu2024prompt}
Y.~Lyu, Y.~Xie, P.~Chen, and H.~Chen, ``Prompt fuzzing for fuzz driver
  generation,'' in \emph{Proceedings of the 2024 on ACM SIGSAC Conference on
  Computer and Communications Security}, 2024, pp. 3793--3807.

\bibitem{liu2025promefuzz}
Y.~Liu, J.~Deng, X.~Jia, Y.~Wang, M.~Wang, L.~Huang, T.~Wei, and P.~Su,
  ``Promefuzz: A knowledge-driven approach to fuzzing harness generation with
  large language models,'' in \emph{Proceedings of the 2025 ACM SIGSAC
  Conference on Computer and Communications Security}, 2025, pp. 1559--1573.

\bibitem{luo2025enhancing}
Z.~Luo, Q.~Du, Y.~Wang, A.~Roychoudhury, and Y.~Jiang, ``Enhancing protocol
  fuzzing via diverse seed corpus generation,'' \emph{IEEE Transactions on
  Software Engineering}, 2025.

\bibitem{ji2026firmagent}
J.~Ji, C.~Zhang, S.~Gan, L.~Jian, H.~Liu, T.~Liu, L.~Zheng, and Z.~Jia,
  ``Firmagent: Leveraging fuzzing to assist llm agents with iot firmware
  vulnerability discovery.'' in \emph{NDSS}, 2026.

\bibitem{sheng2025fuzzingbrain}
Z.~Sheng, Q.~Xu, J.~Huang, M.~Woodcock, H.~Huang, A.~F. Donaldson, G.~Gu, and
  J.~Huang, ``All you need is a {Fuzzing Brain}: An {LLM}-powered system for
  automated vulnerability detection and patching,'' \emph{arXiv preprint
  arXiv:2509.07225}, 2025.

\bibitem{wolff2026large}
D.~Wolff, M.~Mirchev, and A.~Roychoudhury, ``Large language models in software
  security analysis,'' \emph{Communications of the ACM}, vol.~69, no.~6, pp.
  60--67, 2026.

\bibitem{luo2026concollmic}
Z.~Luo, H.~Zhao, D.~Wolff, C.~Cadar, and A.~Roychoudhury, ``Agentic concolic
  execution,'' in \emph{Proceedings of the IEEE Symposium on Security and
  Privacy (S\&P)}, 2026, pp. 1--19.

\bibitem{xia2024fuzz4all}
C.~S. Xia, M.~Paltenghi, J.~Le~Tian, M.~Pradel, and L.~Zhang, ``{Fuzz4All}:
  Universal fuzzing with large language models,'' in \emph{Proceedings of the
  IEEE/ACM 46th International Conference on Software Engineering}, 2024, pp.
  1--13.

\bibitem{du2024vulrag}
X.~Du, G.~Zheng, K.~Wang, Y.~Zou, Y.~Wang, W.~Deng, J.~Feng, M.~Liu, B.~Chen,
  X.~Peng \emph{et~al.}, ``{Vul-RAG}: Enhancing {LLM}-based vulnerability
  detection via knowledge-level {RAG},'' \emph{ACM Transactions on Software
  Engineering and Methodology}, 2024.

\bibitem{park2026agentic}
J.~Park and I.~Yun, ``Agentic fuzzing: Opportunities and challenges,''
  \emph{arXiv preprint arXiv:2605.10074}, 2026.

\bibitem{wang2025vulagent}
Z.~Wang, G.~Li, J.~Li, H.~Zhu, and Z.~Jin, ``{VulAgent}: Hypothesis-validation
  based multi-agent vulnerability detection,'' \emph{arXiv preprint
  arXiv:2509.11523}, 2025.

\bibitem{wei2025smartmultiagent}
Z.~Wei, J.~Sun, Y.~Sun, Y.~Liu, D.~Wu, Z.~Zhang, X.~Zhang, M.~Li, Y.~Liu,
  C.~Li, M.~Wan, J.~Dong, and L.~Zhu, ``Advanced smart contract vulnerability
  detection via {LLM}-powered multi-agent systems,'' \emph{IEEE Transactions on
  Software Engineering}, vol.~51, no.~10, pp. 2830--2846, 2025.

\bibitem{hu2023largesmart}
S.~Hu, T.~Huang, F.~{\.I}lhan, S.~F. Tekin, and L.~Liu, ``Large language
  model-powered smart contract vulnerability detection: New perspectives,'' in
  \emph{2023 5th IEEE International Conference on Trust, Privacy and Security
  in Intelligent Systems and Applications (TPS-ISA)}.\hskip 1em plus 0.5em
  minus 0.4em\relax IEEE, 2023, pp. 297--306.

\bibitem{ernst2007daikon}
M.~D. Ernst, J.~H. Perkins, P.~J. Guo, S.~McCamant, C.~Pacheco, M.~S. Tschantz,
  and C.~Xiao, ``The daikon system for dynamic detection of likely
  invariants,'' \emph{Science of computer programming}, vol.~69, no. 1-3, pp.
  35--45, 2007.

\bibitem{lemieux2015general}
C.~Lemieux, D.~Park, and I.~Beschastnikh, ``General ltl specification mining
  (t),'' in \emph{2015 30th IEEE/ACM international conference on automated
  software engineering (ASE)}.\hskip 1em plus 0.5em minus 0.4em\relax IEEE,
  2015, pp. 81--92.

\bibitem{mu2024clarifygpt}
F.~Mu, L.~Shi, S.~Wang, Z.~Yu, B.~Zhang, C.~Wang, S.~Liu, and Q.~Wang,
  ``Clarifygpt: A framework for enhancing llm-based code generation via
  requirements clarification,'' \emph{Proceedings of the ACM on Software
  Engineering}, vol.~1, no. FSE, pp. 2332--2354, 2024.

\bibitem{endres2024can}
M.~Endres, S.~Fakhoury, S.~Chakraborty, and S.~K. Lahiri, ``Can large language
  models transform natural language intent into formal method postconditions?''
  \emph{Proceedings of the ACM on Software Engineering}, vol.~1, no. FSE, pp.
  1889--1912, 2024.

\bibitem{lahirie2024evaluating}
S.~K. Lahirie, ``Evaluating llm-driven user-intent formalization for
  verification-aware languages,'' in \emph{2024 Formal Methods in
  Computer-Aided Design (FMCAD)}.\hskip 1em plus 0.5em minus 0.4em\relax IEEE,
  2024, pp. 142--147.

\bibitem{zhang2024ecg}
Q.~Zhang, Y.~Shen, J.~Liu, Y.~Xu, H.~Shi, Y.~Jiang, and W.~Chang, ``{ECG}:
  Augmenting embedded operating system fuzzing via {LLM}-based corpus
  generation,'' \emph{IEEE Transactions on Computer-Aided Design of Integrated
  Circuits and Systems}, vol.~43, no.~11, pp. 4238--4249, 2024.

\bibitem{huang2024everything}
H.~Huang, A.~Zhou, M.~Payer, and C.~Zhang, ``Everything is good for something:
  Counterexample-guided directed fuzzing via likely invariant inference,'' in
  \emph{2024 IEEE Symposium on Security and Privacy (SP)}.\hskip 1em plus 0.5em
  minus 0.4em\relax IEEE, 2024, pp. 1956--1973.

\bibitem{zhu2025locus}
\BIBentryALTinterwordspacing
J.~Zhu, C.~Shen, Z.~Li, J.~Yu, Y.~Chen, and K.~Pei, ``Locus: Agentic predicate
  synthesis for directed fuzzing,'' in \emph{Proceedings of the ACM/IEEE 48th
  International Conference on Software Engineering}, ser. ICSE '26.\hskip 1em
  plus 0.5em minus 0.4em\relax New York, NY, USA: Association for Computing
  Machinery, 2026. [Online]. Available:
  \url{https://doi.org/10.1145/3744916.3773102}
\BIBentrySTDinterwordspacing

\bibitem{fioraldi2021use}
A.~Fioraldi, D.~C. D'Elia, and D.~Balzarotti, ``The use of likely invariants as
  feedback for fuzzers,'' in \emph{30th USENIX Security Symposium (USENIX
  Security 21)}, 2021, pp. 2829--2846.

\bibitem{ding2026fmagent}
H.~Ding, Z.~Wang, and H.~Chen, ``{FM-Agent}: Scaling formal methods to large
  systems via {LLM}-based {Hoare}-style reasoning,'' \emph{arXiv preprint
  arXiv:2604.11556}, 2026.

\bibitem{li2024llm}
Z.~Li, S.~Dutta, and M.~Naik, ``Llm-assisted static analysis for detecting
  security vulnerabilities,'' \emph{arXiv preprint arXiv:2405.17238}, 2024.

\bibitem{li2024enhancing}
H.~Li, Y.~Hao, Y.~Zhai, and Z.~Qian, ``Enhancing static analysis for practical
  bug detection: An {LLM}-integrated approach,'' \emph{Proceedings of the ACM
  on Programming Languages}, vol.~8, no. OOPSLA1, pp. 474--499, 2024.

\bibitem{roberts2023cutoff}
M.~Roberts, H.~Thakur, C.~Herlihy, C.~White, and S.~Dooley, ``To the cutoff...
  and beyond? a longitudinal perspective on llm data contamination,'' in
  \emph{The Twelfth International Conference on Learning Representations},
  2023.

\bibitem{deng2024benchmark}
\BIBentryALTinterwordspacing
C.~Deng, Y.~Zhao, X.~Tang, M.~Gerstein, and A.~Cohan, ``Benchmark probing:
  Investigating data leakage in large language models,'' in \emph{NeurIPS 2023
  Workshop on Backdoors in Deep Learning - The Good, the Bad, and the Ugly},
  2024. [Online]. Available: \url{https://openreview.net/forum?id=a34bgvner1}
\BIBentrySTDinterwordspacing

\bibitem{mythos-wolfssl}
\BIBentryALTinterwordspacing
{Anthropic}, ``{ANT-2026-6615Y595}: {wolfSSL} vulnerability finding,'' 2026.
  [Online]. Available:
  \url{https://red.anthropic.com/2026/cvd/findings/ANT-2026-6615Y595}
\BIBentrySTDinterwordspacing

\bibitem{semfix}
H.~Nguyen, D.~Qi, A.~Roychoudhury, and S.~Chandra, ``{SemFix}: Program repair
  via semantic analysis,'' in \emph{2013 35th International Conference on
  Software Engineering (ICSE)}, 2013.

\end{thebibliography}

\end{document}